\documentclass[a4paper,11pt]{article}
\pdfoutput=1
\usepackage{jcappub}

\title{Cosmology with decaying cosmological constant -- exact solutions and model testing}

\author[a,b]{Marek Szyd{\l}owski}

\author[a]{and Aleksander Stachowski}

\affiliation[a]{Astronomical Observatory, Jagiellonian University, Orla 171, 30-244 Krakow, Poland}
\affiliation[b]{Mark Kac Complex Systems Research Centre, Jagiellonian University, ul. {\L}ojasiewicza 11, 30-348 Krak{\'o}w, Poland}

\emailAdd{marek.szydlowski@uj.edu.pl}
\emailAdd{aleksander.stachowski@uj.edu.pl}

\abstract{We study dynamics of $\Lambda(t)$ cosmological models which are a natural generalization of the standard cosmological model (the $\Lambda$CDM model). We consider a class of models: the ones with a prescribed form of $\Lambda(t)=\Lambda_{\text{bare}}+\frac{\alpha^2}{t^2}$. This type of a $\Lambda(t)$ parametrization is motivated by different cosmological approaches. We interpret the model with running Lambda ($\Lambda(t)$) as a special model of an interacting cosmology with the interaction term $-d\Lambda(t)/dt$ in which energy transfer is between dark matter and dark energy sectors. For the $\Lambda(t)$ cosmology with a prescribed form of $\Lambda(t)$ we have found the exact solution in the form of Bessel functions. Our model shows that fractional density of dark energy $\Omega_e$ is constant and close to zero during the early evolution of the universe.

We have also constrained the model parameters for this class of models using the astronomical data such as SNIa data, BAO, CMB, measurements of $H(z)$ and the Alcock-Paczy{\'n}ski test. In this context we formulate a simple criterion of variability of $\Lambda$ with respect to $t$ in terms of variability of the jerk or sign of estimator $(1-\Omega_{\text{m},0}-\Omega_{\Lambda,0})$. The case study of our model enable us to find an upper limit $\alpha^2 < 0.012$ ($2\sigma$ C.L.) describing the variation from the cosmological constant while the LCDM model seems to be consistent with various data.}

\begin{document}
\maketitle
\flushbottom

\section{Introduction}
The standard cosmological model describes the matter content of the Universe comprising the cold dust matter (baryonic matter and dark matter) which satisfies the equation of state for dust $p=0$. In turn, dark energy is described in terms of an effective parameter (the cosmological constant) which should be treated as the best `economical' (as only one parameter used to describe the whole dark sector) description of the cause that the Universe expansion accelerates in the current epoch.

The natural interpretation of the cosmological constant arises as an effect of quantum vacuum energy. Since this form of energy should be independent of the reference frame it must be proportional to the only `invariant' second order metric tensor $g_{\mu \nu}$, i.e. $T_{\mu \nu}=\rho_{\text{vac}} g_{\mu \nu}$. If we include the conservation condition which for the cosmological model with the Robertson-Walker (R-W) symmetry assumes the form
\begin{equation} \label{eq:1}
    \dot\rho=-3H(\rho+p),
\end{equation}
then we obtain that $\rho_{\text{vac}}=\text{const}=\Lambda$ and $p_{\text{vac}}=-\Lambda$, $H=\frac{d}{dt}(\ln a)$ is the Hubble parameter, where $\rho$ is total energy density, $p$ is total pressure, an overdot denotes differentiation with respect to the cosmological time $t$; we use a natural system of units in which $8\pi G=c=1$.

If we interpret the cosmological constant $\Lambda$ as a vacuum energy, then there is a difference between its today value required to explain observations of type Ia supernovae (SNIa) and the value of $\rho_{\text{vac}}$ estimated from effective field theory. The former is smaller by a factor of $10^{-120}$. This discrepancy is called the cosmological constant problem.

To achieve the conservation of energy-momentum tensor (divergence of energy-momentum tensor $T_{\mu \nu}$ is vanishing) different descriptions of dark energy sector have been proposed. In the simplest case the time cosmological term $\Lambda(t)$ is shifted to the right-hand side and treated as a source of gravity. Such an approach is called a $\Lambda(t)$CDM cosmology.

In this paper we assume $\rho=\rho_{\text{m}}+\rho_{\text{de}}$ where $\rho_{\text{m}}$ is a density of matter and $\rho_{\text{de}}$ is the density of dark energy. We also assume $\rho_{\text{m}}=\rho_{\text{b}}+\rho_{\text{dm}}$, and $p_{\text{m}}=p_{\text{b}}+p_{\text{dm}}$, where $\rho_{\text{b}}=\rho_{\text{b,0}} a(t)^{-3}$ and $p_{\text{b}}=0$ are a density and a pressure of baryonic matter, $\rho_{\text{dm}}$ and $p_{\text{dm}}=0$ are a density and a pressure of dark matter, $a(t)$ is the scale factor. The state equation for dark energy is assumed as $p_{\text{de}}=-\rho_{\text{de}}$. In this case the conservation condition has the following form
\begin{align} \label{eq:2}
    \dot\rho_{\text{dm}}+3H\rho_{\text{dm}} &= Q,\\
    \dot\rho_{\text{b}}+3H\rho_{\text{b}} &= 0
\end{align}
and
\begin{equation} \label{eq:3}
    \dot\rho_{\text{de}}=-Q,
\end{equation}
where $Q$ describes an interaction between dark matter and dark energy and this case is expressed by $Q=-\dot\Lambda$.
The conservation condition can be rewritten in the form
\begin{equation} \label{eq:4}
    \dot\rho_{\text{m}}=-3H\rho_{\text{m}}-\dot\Lambda,
\end{equation}
where $\rho_{\text{m}}$ and $p_{\text{m}}$ are energy density and pressure of matter.

Pani et al. considered the energy-momentum tensor which ensures the covariantness of general relativity \cite{Pani:2013qfa}. An alternative approach is to postulate the scalar field $\phi$ with the potential $V(\phi)$ for this model \cite{Szydlowski:2015bwa} which guarantee that model is covariance.

We consider a model with a parametrization of $\Lambda$ following the rule
\begin{equation} \label{eq:5}
    \Lambda(t)=\Lambda_{\text{bare}}+\frac{\alpha^2}{t^2},
\end{equation}
where $\alpha^2$ is a real constant; $\Lambda_{\text{bare}}$ is a constant and $\rho_{\text{vac}}=\Lambda$. This model belongs to a larger class of cosmological models with interaction. In this case the interaction term is $Q=-d\Lambda/dt$. In this model the interaction is between dark matter and dark energy. This model belongs to a class of models so-called early constant dark energy during the matter dominating stage.

If we replace the cosmological time $t$ by the Hubble scale time in eq.~(\ref{eq:5}), then we obtain the $\Lambda(H)$ parametrization which is based on Lima at al. \cite{Lima:1995kd,Lima:2015kda,Perico:2013mna}.

We estimate the value of the parameter $\alpha^2$ as well as the other models parameters from available astronomical data. This class of models is compared with the standard cosmological model (the $\Lambda$CDM model).

Let us enumerate motivations for introducing form (\ref{eq:5}) of parametrization of dark energy.
\begin{enumerate}
    \item The parametrization of dark energy can be derived from the quantum mechanics which describes how decaying false vacuum states changes in time. It can be shown that at the late time it can be identified as the cosmological constant which is time dependent and changes following the rule (\ref{eq:5}) and parameter $\alpha^2$ is small and constitutes a leading term for long-term behaviour in power series of energy density of decaying vacuum  \cite{Szydlowski:2015kqa,Szydlowski:2015bwa}\cite{Urbanowski:2013tfa,Urbanowski:2012pka}.
    \item A new model of agegraphic dark energy \cite{Cai:2007us,Wei:2007ty} based on some quantum arguments that the energy density of metric fluctuation of the Minkowski spacetime is proportional to $\frac{1}{t^2}$ and and it also motivated K{\'a}rolyh{\'a}zy uncertainty relation \cite{Maziashvili:2007dk}. If we identify the time scale as the age of the Universe $T$, then we obtain that the agegraphic dark energy is $\rho_q\propto\frac{1}{T^2}$.
    \item In the de Sitter universe there is a possibility to define in the framework of general relativity length and time scales $\Lambda(t)=\frac{3}{r_{\Lambda}^2 (t)}=\frac{3}{c^2 t_{\Lambda}^2 (t)}$ \cite{Chen:2011rz}. Otherwise, any cosmological length scale or time scale can determined the relation $\Lambda(t)$. Chen et al. \cite{Chen:2011rz} demonstrated how holographic \cite{Li:2004rb,Zhang:2012pr} and agegraphic dark energy conceptions can be unified in the framework of interacting cosmology in which the interacting term is $Q=-\dot\rho_{\Lambda}$. The variational approach to an interacting quintessence model was recently considered by B{\"o}hmer et al. \cite{Boehmer:2015kta}.
    \item Ringermacher and Mead \cite{Ringermacher:2014kaa} considered dark matter as a perfect fluid satisfying the equation of state $p=-\frac{1}{3}\rho$. The energy density of such fluid mimicking dark matter effects varies like $\frac{1}{t^2}$ rather than $\frac{\Omega_{\text{dark}}}{a^3}$ as in the standard cosmological model.
    \item Haba has discussed recently cosmological models of general relativity in which a source of gravity (right-hand sides of the Einstein equations) is a sum of the energy-momentum of particles and the cosmological term describing a dissipation of energy-momentum. He obtained a cosmological model with the cosmological term decaying as $1/t^2$ \cite{Haba:2013xka,Haba:2014fd}.
\end{enumerate}

\section{Exact solutions for $\Lambda(t)$CDM cosmology with $\Lambda(t)=\Lambda+\frac{\alpha^2}{t^2}$}

For the parametrization of $\Lambda(t)$ (\ref{eq:5}) it is possible to obtain exact solutions and discuss cosmological implications of this generalized standard cosmological model. We show that a deviation of this model from the $\Lambda$CDM model can be probed by a measurement of a jerk.

We start from the Friedmann first integral in the FRW cosmology with $\Lambda(t)=\Lambda_{\text{bare}}+\frac{\alpha^2}{t^2}$, where $t$ is the cosmological time and $\alpha^2$ is either positive or negative,
\begin{equation}
    3H(t)^2=\rho_{\text{m}}(t)+\Lambda_{\text{bare}}+\frac{\alpha^2}{t^2}\label{friedmann}
\end{equation}
and the conservation condition
\begin{equation}
    \dot\rho_{\text{m}}(t)=-3H(t)\rho_{\text{m}}(t)-\frac{d(\Lambda_{\text{bare}}+\frac{\alpha^2}{t^2})}{d t}.\label{conservation}
\end{equation}
Equation (\ref{friedmann}) can be rewritten in the dimensionless parameters
\begin{equation}
    \Omega_{\text{m},0}=\frac{\rho_{\text{m,0}}}{3H_0^2}, \quad \Omega_{\Lambda,0}=\frac{\Lambda_{\text{bare}}}{3H_0^2}, \quad \Omega_{\alpha,0}=\frac{\alpha^2}{3H_0^2 T^{2}_{0}},
\end{equation}
where $T_0$ is the present age of the Universe, i.e. $T_0=\int_0^{T_{0}} dt=\int_0^{a_0} \frac{da}{Ha}$, and quantities labeled by index `0' are defined at the present epoch for which $a_0=1$. Then equation (\ref{friedmann}) has the following form
\begin{equation}
    \frac{H(t)^2}{H_0^2}=\Omega_{\text{m}}(t)+\Omega_{\Lambda,0}+\Omega_{\alpha,0}\frac{T_0^2}{t^2}, \label{condition}
\end{equation}
where $\Omega_{\text{m}}(t)=\Omega_{\text{b,0}}a(t)^{-3}+\Omega_{\text{dm,0}}f(t)$ and $f(T_0)=1$.
At present, equation (\ref{condition}) is expressed by
\begin{equation}
    1 = \Omega_{\text{m},0} + \Omega_{\Lambda,0} + \Omega_{\alpha,0}.
\end{equation}
After differentiation of both sides of (\ref{friedmann}) with respect to $t$ we obtain
\begin{equation} \label{eq:10}
    6H(t)\dot H(t)=\dot\rho_{\text{m}} (t)+\frac{d(\Lambda_{\text{bare}}+\frac{\alpha^2}{t^2})}{d t}.
\end{equation}
Equation (\ref{eq:10}) can be simplified with the help of (\ref{conservation}). Then we obtain
\begin{equation}
    \dot H(t)=-\frac{\rho_{\text{m}}(t)}{2}.\label{dec}
\end{equation}
After substitution of (\ref{friedmann}) to (\ref{dec}) we obtain
\begin{equation}
    \dot H(t)=\frac{1}{2}\left(\Lambda_{\text{bare}}+\frac{\alpha^2}{t^2}-3H(t)^2\right)\label{eq1}.
\end{equation}
Equation (\ref{eq1}) can be rewritten in the dimensionless parameters. Then we obtain
\begin{equation}
    \dot h(t)=\frac{3H_0}{2}\left(\Omega_{\Lambda,0}+\frac{\Omega_{\alpha,0}T^{2}_{0}}{t^2}-h(t)^2\right),\label{eq2}
\end{equation}
where $h(t)=\frac{H(t)}{H_0}$.

The general solution of equation (\ref{eq2}) has the following form
\begin{equation}
    h(t)=\frac{2}{3 H_0} \frac{d}{dt} \log\left[\sqrt t \left(C_1 Y_n \left(\frac{3\sqrt {-\Omega_{\Lambda,0}} H_0}{2}t\right)+
J_n \left(\frac{3\sqrt {-\Omega_{\Lambda,0}} H_0}{2}t \right)\right)\right],\label{eq:15}
\end{equation}
where $C_1$ is a constant, an $J_n(x)$ and $Y_n(x)$ are Bessel functions of the first and second kind, the index $n$ of these functions is given in terms of $\Omega_{\alpha,0}$, $H_0$ and $T_0$, $n=\frac{1}{2}\sqrt{1+9\Omega_{\alpha,0}T^2_0 H_0^2}$. We can rewrite (\ref{eq:15}) to the form
\begin{equation}
    h(t)=\frac{2}{3 H_0} \frac{d}{dt} \log\left[\sqrt t \left(D_1 Y_n \left(\frac{3\sqrt {-\Omega_{\Lambda,0}} H_0}{2}t\right)+
D_2 J_n \left(\frac{3\sqrt {-\Omega_{\Lambda,0}} H_0}{2}t \right)\right)\right],\label{sol}
\end{equation}
For the correspondence with the $\Lambda$CDM model ($\alpha^2=0$) we choose $D_1=0$. Then solution (\ref{sol}) is given by the formula
\begin{equation}
    h(t)=\frac{2}{3 H_0} \frac{d}{dt} \log\left[\sqrt t \left(
I_n \left(\frac{3\sqrt {\Omega_{\Lambda,0}} H_0}{2}t \right)\right)\right],\label{sol2}
\end{equation}
where $I_n(x)$ is the modified Bessel function. Solution (\ref{sol2}) can be rewritten to the following form
\begin{equation}
    h(t)=\frac{1-2n}{3H_0 t}+\sqrt{\Omega_{\Lambda,0}}\frac{I_{n-1}\left(\frac{3\sqrt {\Omega_{\Lambda,0}} H_0}{2}t \right)}{
I_n \left(\frac{3\sqrt {\Omega_{\Lambda,0}} H_0}{2}t \right)}.
\end{equation}

\begin{figure}[ht]
\centering
\includegraphics[scale=0.85]{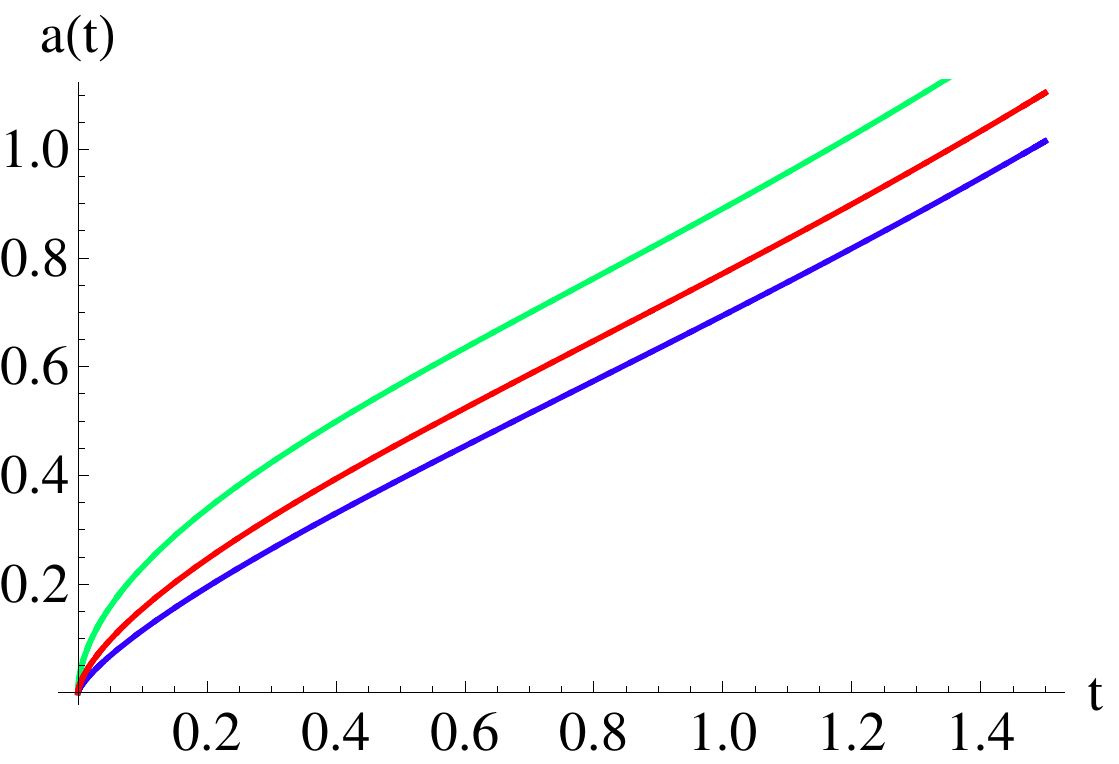}
\caption{Diagram of the scale factor $a(t)$ for three cases. The top function is for $\alpha^2=-0.2$, the middle function represents the $\Lambda$CDM model and the bottom function is for $\alpha^2=0.2$. We assume that $H_0=68.27$ km/(s Mpc) and $\Omega_{\text{m}}=0.35$. Time $t$ is expressed in a unit $(\text{100 s Mpc/km})$.}
\label{fig:1}
\end{figure}

\begin{figure}[ht]
\centering
\includegraphics[scale=0.85]{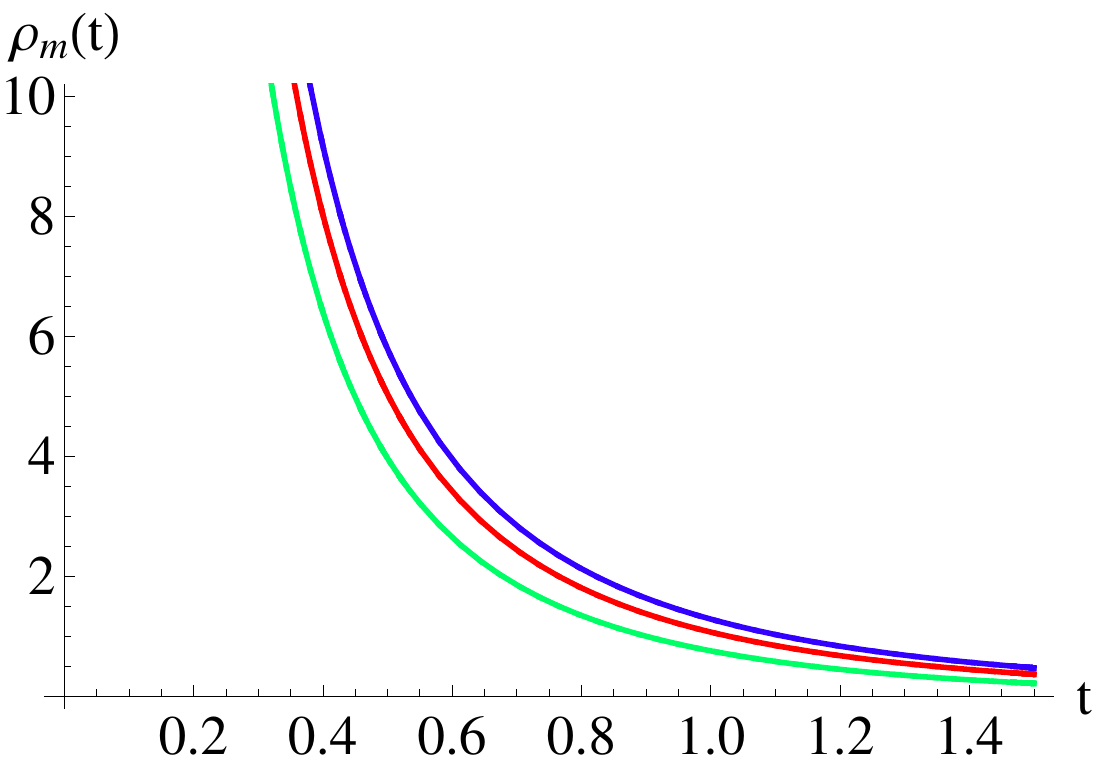}
\caption{Diagram of the energy density $\rho_{\text{m}}(t)$ for three cases. The top function is for $\alpha^2=0.2$, the middle function represents the $\Lambda$CDM model and the bottom function is for $\alpha^2=-0.2$. We assume that $H_0=68.27$km/(s Mpc) and $\Omega_{\text{m,0}}=0.35$. Time t is expressed in unit $(\text{100 s Mpc/km})$. We assume $8\pi G=1$ and we choose for $\rho_{\text{m}}$ a unit (km/(100 s Mpc))$^2$.}
\label{fig:2}
\end{figure}

\begin{figure}[ht]
\centering
\includegraphics[scale=0.85]{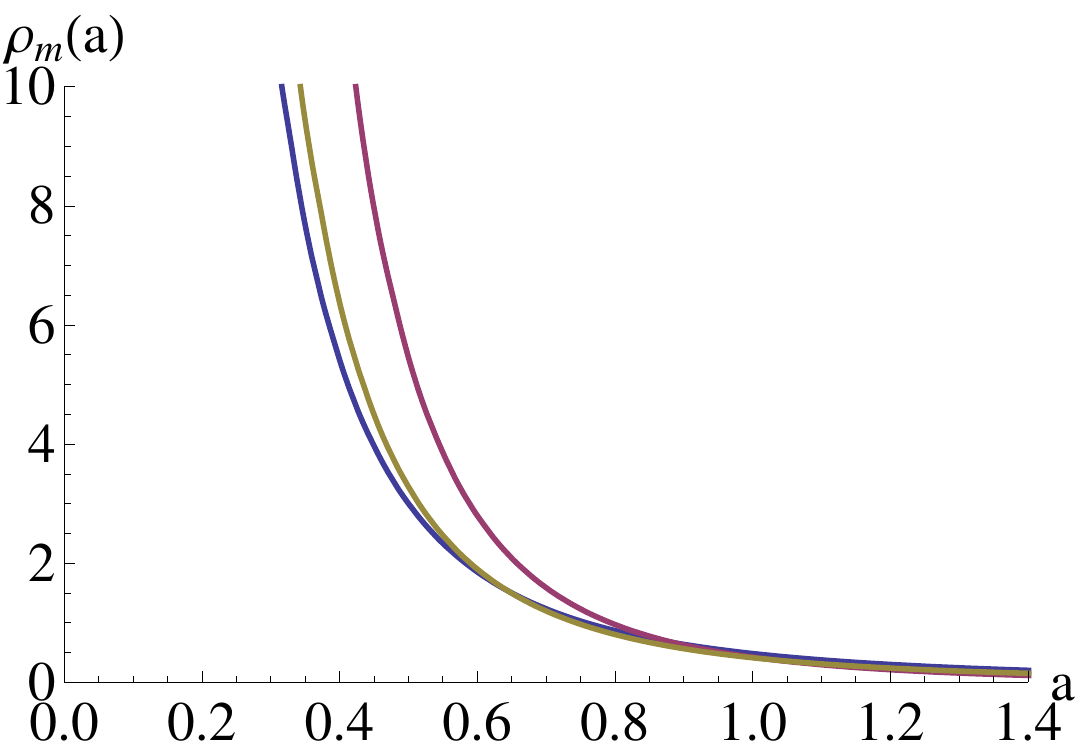}
\caption{Diagram of the energy density $\rho_{\text{m}}(a)$ for three cases. The top function is for $\alpha^2=-0.2$, the middle function represents the $\Lambda$CDM model and the bottom function is for $\alpha^2=0.2$. We assume that $H_0=68.27$km/(s Mpc) and $\Omega_{\text{m,0}}=0.35$. We assume $8\pi G=1$ and we choose for $\rho_{\text{m}}$ a unit (km/(100 s Mpc))$^2$.}
\label{fig:10}
\end{figure}

\begin{figure}[ht]
\centering
\includegraphics[scale=0.85]{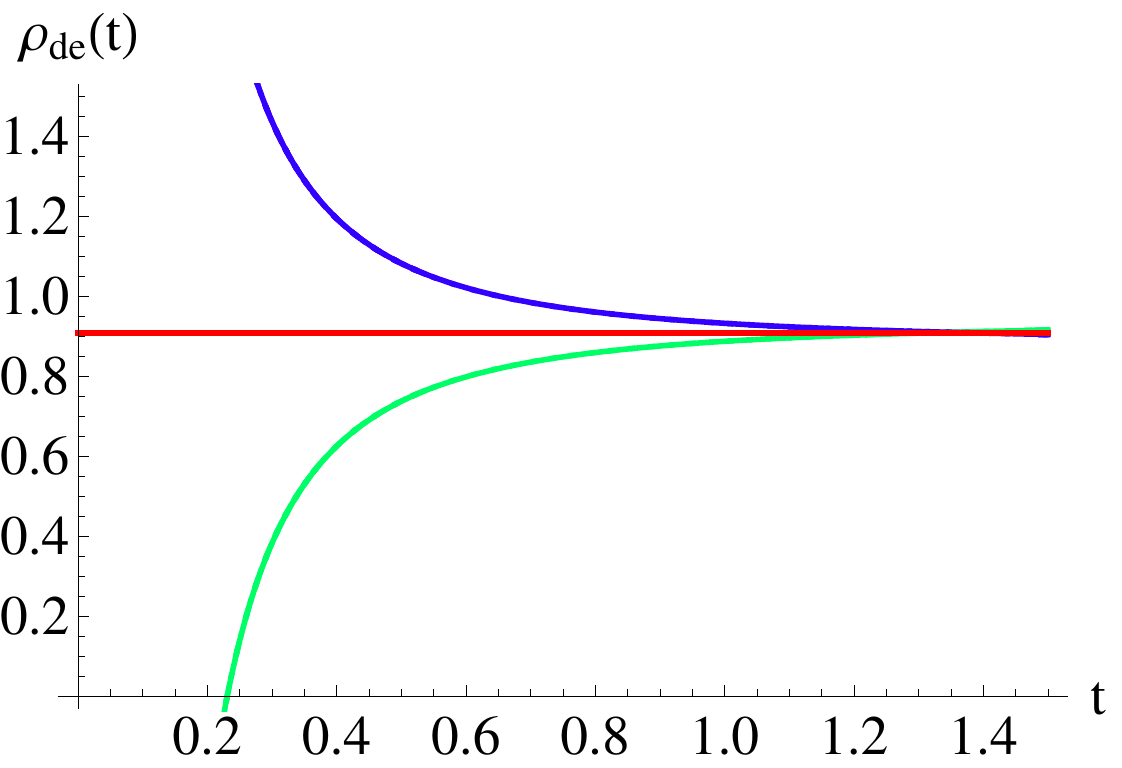}
\caption{Diagram of the energy density $\rho_{\text{de}}(t)$ for three cases. The top function is for $\alpha^2=0.05$, the middle function represents the $\Lambda$CDM model and the bottom function is for $\alpha^2=-0.05$. We assume that $H_0=68.27$km/(s Mpc) and $\Omega_{\text{m,0}}=0.35$. Time $t$ is expressed in unit $(\text{100 s Mpc/km})$. We assume $8\pi G=1$ and we choose for $\rho_{\text{de}}$ a unit (km/(100 s Mpc))$^2$.}
\label{fig:14}
\end{figure}

Because $H(t)=\frac{d}{dt}\ln a$, then it is easy to obtain the scale factor from (\ref{sol2}) in the form
\begin{equation}
     a(t)=C_2 \left[\sqrt t \left(
I_n \left(\frac{3\sqrt {\Omega_{\Lambda,0} }H_0}{2}t\right)\right)\right]^\frac{2}{3}.\label{factor}
\end{equation}
The diagram of $a(t)$ is presented in figure \ref{fig:1}. We obtain a formula for $\rho_{\text{m}} (t)$ from (\ref{dec}), (\ref{eq2}) and (\ref{sol2})
\begin{equation}
    \rho_{\text{m}}=-3H_0^2\left(\Omega_{\Lambda,0}+\frac{\Omega_{\alpha,0}T^2_0}{t^2}-\left(\frac{2}{3 H_0} \frac{d}{dt} \log\left[\sqrt t \left(
I_n \left(\frac{3\sqrt {\Omega_{\Lambda,0}} H_0}{2}t \right)\right)\right]\right)^2\right)
\end{equation}
or in an equivalent form
\begin{equation}
    \rho_{\text{m}}=-3H_0^2\left(\Omega_{\Lambda,0}+\frac{\Omega_{\alpha,0}T^2_0}{t^2}-\left(\frac{1-2n}{3H_0 t}+\sqrt{\Omega_{\Lambda,0}}\frac{I_{n-1}\left(\frac{3\sqrt {\Omega_{\Lambda,0}} H_0}{2}t \right)}{
I_n \left(\frac{3\sqrt {\Omega_{\Lambda,0}} H_0}{2}t \right)}\right)^2\right).\label{matter}
\end{equation}
The diagrams of $\rho_{\text{m}} (t)$ and $\rho_{\text{m}} (a)$ are presented in figure \ref{fig:2} and \ref{fig:10}. In comparison, the diagram of $\rho_{\text{de}} (t)$ is demonstrated in figure \ref{fig:14}.

The dark matter is expressed by
\begin{equation}
    \rho_{\text{dm}}=\rho_{\text{m}}-\rho_{\text{b},0}a^{-3}.\label{matter2}
\end{equation}
If we use formulas (\ref{factor}) and (\ref{matter}) in equation~(\ref{matter2}) then we get
\begin{multline}
    \rho_{\text{dm}}=-3H_0^2\left(\Omega_{\Lambda,0}+\frac{\Omega_{\alpha,0}T^2_0}{t^2}-\left(\frac{1-2n}{3H_0 t}+\sqrt{\Omega_{\Lambda,0}}\frac{I_{n-1}\left(\frac{3\sqrt {\Omega_{\Lambda,0}} H_0}{2}t \right)}{
I_n \left(\frac{3\sqrt {\Omega_{\Lambda,0}} H_0}{2}t \right)}\right)^2\right) \\
-\rho_{\text{b,0}} C_2^{-3} \left[\sqrt t \left(
I_n \left(\frac{3\sqrt {\Omega_{\Lambda,0} }H_0}{2}t\right)\right)\right]^{-2}.
\end{multline}

The $\Lambda$CDM model can be obtained in the limit $\Omega_{\alpha,0}=0$. Then index $n=\frac{1}{2}$ and $I_{\frac{1}{2}}(x)=\sqrt\frac{2}{\pi x}\sinh(x)$. Finally the solution (\ref{sol2}) reduces to
\begin{equation} \label{eq:23}
    h(t)=\frac{2}{3 H_0} \frac{d}{dt} \log\left[
\sinh\left(\frac{3\sqrt {\Omega_{\Lambda,0}} H_0}{2}t \right)\right].
\end{equation}
Equation (\ref{eq:23}) can be rewritten to the equivalent form
\begin{equation}
    h(t)=\sqrt {\Omega_{\Lambda,0}}
\coth\left(\frac{3\sqrt {\Omega_{\Lambda,0}} H_0}{2}t \right).
\end{equation}
In the special case the solution of (\ref{eq2}) for $\Omega_{\Lambda,0}=0$ has the following form
\begin{equation} \label{eq:25}
    h(t)=\frac{1}{3Ht}\left[1+\sqrt{1+9H^2_0\Omega_{\alpha,0}T^2_0}\left(1-\frac{C_1}{t^{\sqrt{1+9H^2_0\Omega_{\alpha,0}T^2_0}}+C_1}\right)\right].
\end{equation}
For the correspondence with the CDM model we choose $C_1=0$. Then equation (\ref{eq:25}) is simplified to
\begin{equation}
    h(t)=\frac{1}{3H_0 t}\left[1+\sqrt{1+9H^2_0 \Omega_{\alpha,0}T^2_0}\right].\label{cdm}
\end{equation}
From equation (\ref{cdm}) we can obtain an expression for the scale factor
\begin{equation}
    a(t)=C_2 t^{\frac{1}{3}\left(1+\sqrt{1+9H^2_0 \Omega_{\alpha,0}T^2_0}\right)}.
\end{equation}

\begin{figure}[ht]
\centering
\includegraphics[scale=0.85]{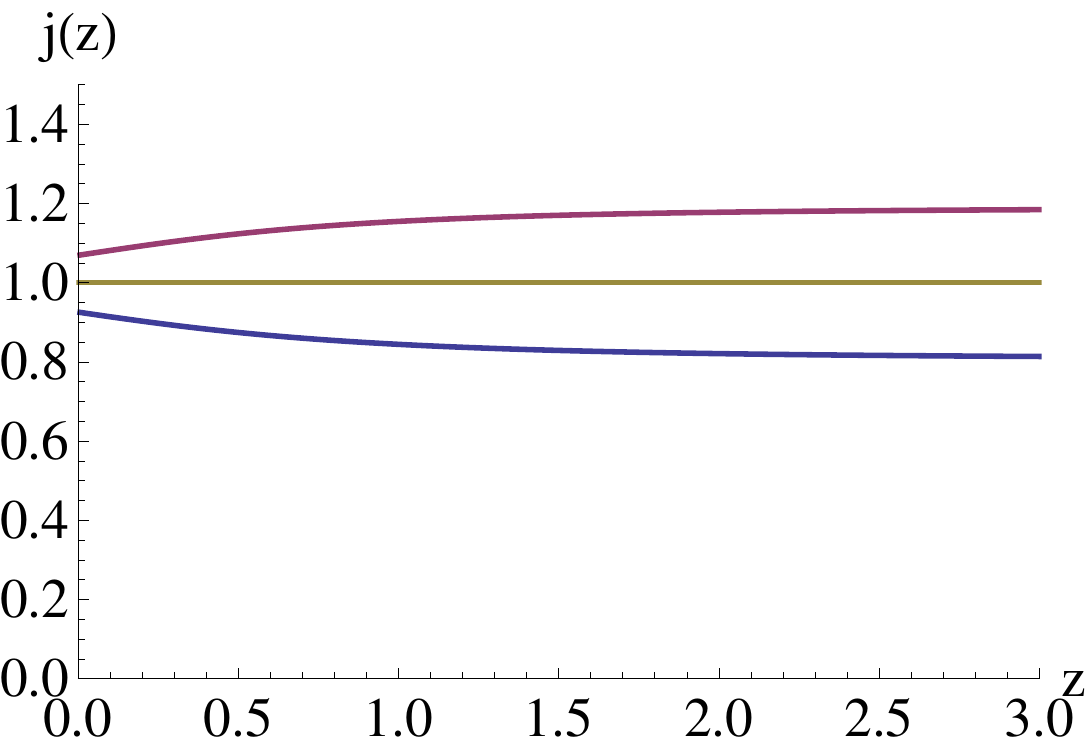}
\caption{Diagram of $j(z)$ for $H_0=68.27$ km/(s Mpc) and $\Omega_{\text{m},0}=0.35$. The top function is for $\alpha^2=-0.05$, the middle function represents the $\Lambda$CDM model and the bottom function is for $\alpha^2=0.05$.}
\label{fig:12}
\end{figure}

\begin{figure}[ht]
\centering
\includegraphics[scale=0.85]{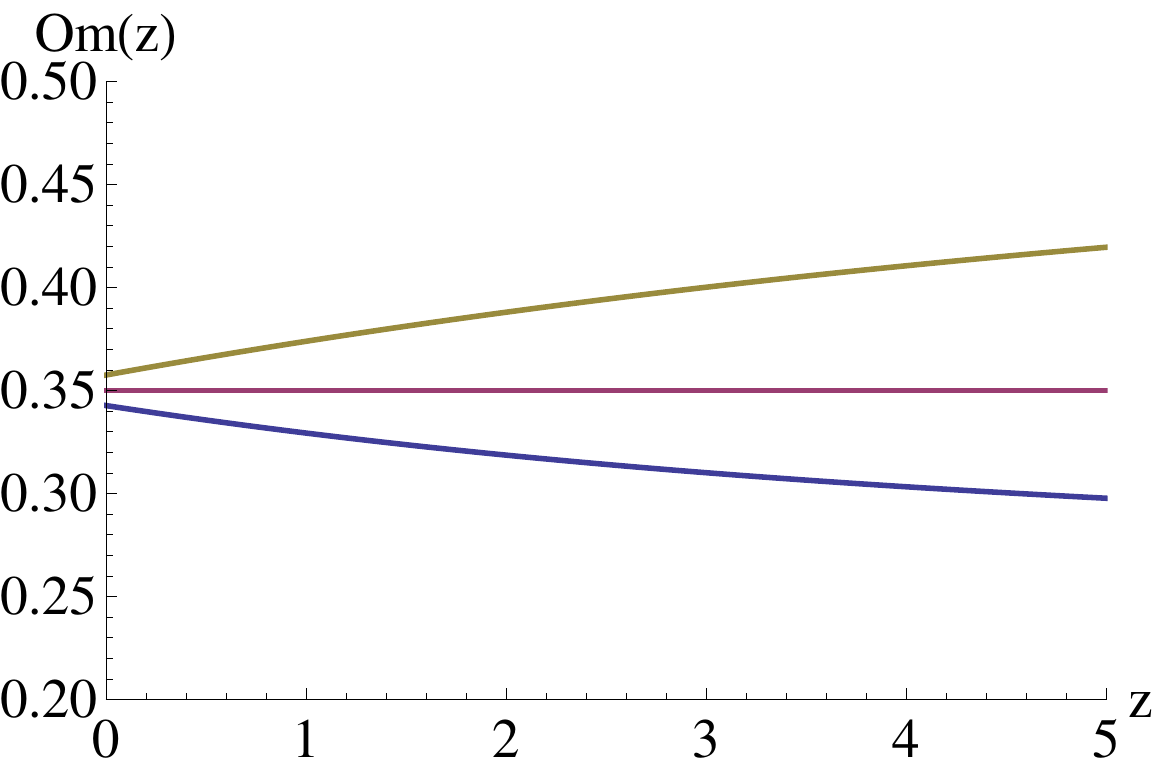}
\caption{Diagram of $Om(z)$ for $H_0=68.27$ km/(s Mpc) and $\Omega_{\text{m},0}=0.35$. The top function is for $\alpha^2=-0.05$, the middle function represents the $\Lambda$CDM model and the bottom function is for $\alpha^2=0.05$.}
\label{fig:11}
\end{figure}

If we know an exact solution for the scale factor $a(t)$ it will be possible to calculate a dimensionless parameter called a jerk related with a third order time derivative of the scale factor
\begin{equation}
    j = \frac{1}{H(t)^3 a(t)}\left[\frac{d^3a(t)}{dt^3}\right].
\end{equation}
After some calculations we obtain the third order time derivative of the scale factor in the form
\begin{equation}
    \dddot a=\frac{3H_0^2 \dot a}{2}\left(\Omega_{\Lambda,0}+\frac{\Omega_{\alpha,0}T^2_0}{t^2}-h^2\right)+\frac{3H_0^2 a}{2}\left(-\frac{2\Omega_{\alpha,0}T^2_0}{t^3}-2h\dot h\right)+H_0^2 \dot a h^2+2H_0^2 a h\dot h.
\end{equation}
A substitution of the expression $h(t)$ from (\ref{eq2}) gives us the exact formula for the jerk as a function of the cosmological time $t$
\begin{equation}
    j(t)=1- \frac{3\Omega_{\alpha,0}T^2_0}{H_0 t^3} \left(\frac{1-2n}{3H_0 t}+\sqrt{\Omega_{\Lambda,0}}\frac{I_{n-1}\left(\frac{3\sqrt {\Omega_{\Lambda,0}} H_0}{2}t \right)}{
I_n \left(\frac{3\sqrt {\Omega_{\Lambda,0}} H_0}{2}t \right)}\right)^{-3}.\label{jerk}
\end{equation}
The jerk calculated for $t=T_0$, i.e. for the present epoch is given by formula
\begin{equation}
    j_0 = 1- \frac{3 \Omega_{\alpha,0}}{H_0 T_0}.\label{jerk2}
\end{equation}
The diagram of $j(z)$ is presented in figure~\ref{fig:12}. One can see the jerk can be treated as a tool for detection the variability of dark energy.

From the exact solution (\ref{jerk}) one can see that the deviation of the generalized model from the $\Lambda$CDM model is given by time dependent contribution to the jerk because for the $\Lambda$CDM model it is equal one. Therefore if we can detect from the astronomical observations the time variability of the jerk it will be a simple diagnostic of decaying vacuum. If $\Omega_{\alpha,0}$ is non-zero this means that $\Omega_{\text{m},0}+\Omega_{\Lambda,0}<1$. Note that
\begin{equation}
    \Omega_{\alpha,0}=1-\Omega_{\text{m},0}-\Omega_{\Lambda,0}=\frac{\alpha^2}{3H_0^2 T^2_0}.
\end{equation}
Because $T_0\leq\frac{1}{H_0}$, i.e. $H_0^2 T^2_0\leq 1$ and $\alpha^2=3H_0^2 T^2_0 \Omega_{\alpha,0}$, i.e.
\begin{equation}
    \frac{\alpha^2}{3}\leq\Omega_{\alpha,0}
\end{equation}
and from the estimation of $\Omega_{\alpha,0}$ one can obtain an upper limit on $\frac{\alpha^2}{3}$.

From the value of the jerk for the current epoch (see formula (\ref{jerk2})) there comes also the limit values of the jerk
\begin{equation}
    1-3\Omega_{\alpha,0}\leq j_0\leq 1, \qquad \text{for } \alpha^2 > 0
\end{equation}
and
\begin{equation}
1 \leq j_0 \leq 1 - 3 \Omega_{\alpha,0}, \qquad \text{for } \alpha^2 < 0.
\end{equation}

Sahni et al. \cite{Sahni:2008xx,Shafieloo:2012rs,Sahni:2014ooa} proposed in the context of testing and comparison of alternatives for the $\Lambda$CDM model $Om(z)$ diagnostic test
\begin{equation}
    Om(z)=\frac{h^2(x)-1}{x^3-1},
\end{equation}
where $x=1+z$. While this parameter is constant for the $\Lambda$CDM model, $Om(x)=\Omega_{\text{m},0}$ for any deviation from zero would discard the $\Lambda$CDM model for the description of the cosmic evolution of the current Universe for low $z$. But Om($z$) diagnostic test is not constant for high $z$ because $\Lambda$CDM model should respect radiation for high $z$. Note that if the radiation density is included then the behavior of $Om(z)$ will be different for the case of matter and cosmological constant. For high redshift the contribution from radiation density will dominate. In our paper matter and energy density is present at very beginning and effect of radiation density is not included because of complexity of analytical calculations. Therefore our comparison of a jerk and $Om(z)$ is not valid for high redshift. Let us note that in our case $Om(x)$ is not constant and evolves with the cosmological time as
\begin{equation}
    Om(t)=\frac{\left(\frac{1-2n}{3H_0 t}+\sqrt{\Omega_{\Lambda,0}}\frac{I_{n-1}\left(\frac{3\sqrt {\Omega_{\Lambda,0}} H_0}{2}t \right)}{
I_n \left(\frac{3\sqrt {\Omega_{\Lambda,0}} H_0}{2}t \right)}\right)^2-1}{\left(\left[\sqrt T_0 \left(
I_n \left(\frac{3\sqrt {\Omega_{\Lambda,0} }H_0}{2}T_0 \right)\right)\right]^{2} \left[\sqrt t \left(
I_n \left(\frac{3\sqrt {\Omega_{\Lambda,0} }H_0}{2}t\right)\right)\right]^{-2}\right)-1}.
\end{equation}
The diagram of $Om(z)$ is presented in figure~\ref{fig:11}. From comparison of figures~\ref{fig:12} and \ref{fig:11} one can observe two alternative ways of the detection of the variability of dark energy with respect to time.

\section{Dynamics of the generalized $\Lambda$CDM model}

For a deeper analysis of dynamics it is useful to investigate how exact solutions (trajectories) depend on initial conditions. The natural language for such a discussion is the phase space which a space of all solutions for all admissible initial conditions.

Let us consider now the dynamics of the model under consideration as a dynamical system. In this paper we consider the case of a positive cosmological constant $\Lambda_{\text{bare}}>0$ and strictly positive energy density of matter $\rho_{\text{m}}>0$. The first step in a formulation of the dynamics in terms of a dynamical system is a choice of the state variables. Assume state variables are as follows
\begin{equation} \label{eq:35}
x^2=\frac{\rho_{\text{m}}}{3H^2}, \quad y^2=\frac{\Lambda_{\text{bare}}}{3H^2}, \quad z^2=\frac{1}{3H^2 t^2}.
\end{equation}
We also choose a new time variable $\tau \colon \tau=\ln a$; let a prime denotes the differentiation with respect to the Hubble time $\tau$. Then, we differentiate with respect to $\tau$ the expressions for $x^2$, $y^2$ and $z^2$ in (\ref{eq:35}) and obtain
\begin{align}
    2xx' &= \frac{2x\dot x}{H}=\frac{\dot\rho_{\text{m}}}{3H^3}-\frac{2\rho_{\text{m}} \dot H}{3H^4}, \\
    2yy' &= \frac{2y\dot y}{H}=-\frac{2\Lambda_{\text{bare}} \dot H}{3H^4}, \\
    2zz' &= \frac{2z\dot z}{H}=-\frac{2}{3H^3 t^3}-\frac{2 \dot H}{3H^4 t^2}.
\end{align}
Due to relation (\ref{conservation}) the expression for $\dot\rho_{\text{m}}$ can be replaced by $-3H\rho_{\text{m}} +\frac{2\alpha^2}{t^3}$ and then with the help of (\ref{dec}) $\dot H$ can be replaced by $-\frac{\rho_{\text{m}}}{2}$. As a consequence we obtain the set of equations
\begin{align}
    2xx' &= -\frac{\rho_{\text{m}}}{H^2}+\frac{2\alpha^2}{3H^3 t^3}+\frac{2\rho_{\text{m}}^{2}}{6H^4}, \\
    2yy' &= \frac{2\Lambda_{\text{bare}} \rho_{\text{m}}}{6H^4}, \\
    2zz' &= -\frac{2}{3H^3 t^3}+\frac{2 \rho_{\text{m}}}{6H^4 t^2}.
\end{align}
After returning to the original variables $x$, $y$, $z$ we obtain the system
\begin{align}
    x' &= -\frac{3}{2}x+\sqrt{3}\alpha^2 \frac{z^3}{x}+\frac{3}{2}x^3,\label{dyn1} \\
    y' &= \frac{3}{2}x^2 y,\label{dyn2} \\
    z' &= -\sqrt{3}z^2+\frac{3}{2}x^2 z.\label{dyn3}
\end{align}
Note that the right-hand side of (\ref{dyn1}) is not defined on the plane $x=0$. All state variables are constrained by the condition $x^2+y^2+z^2=1$, i.e. phase space is a surface of a three-dimensional sphere.

We regularize system (\ref{dyn1})-(\ref{dyn3}) in such a way that its right-hand sides are in a polynomial form. For this purpose we introduce new state variables $X, Y, Z \colon X=x^2, Y=y, Z=z$. Note that transformation $x\rightarrow X$ is not a diffeomorphism on the line $x=0$. Then system (\ref{dyn1})-(\ref{dyn3}) represents the dynamical system with smooth right-hand side functions, namely
\begin{align}
    X' &= -3X+{3}X^2+2\sqrt{3}\alpha^2 Z^3, \label{dynX} \\
    Y' &= \frac{3}{2}X Y, \label{dynY} \\
    Z' &= -\sqrt{3}Z^2+\frac{3}{2}Z X, \label{dynZ}
\end{align}
where the phase space is restricted by the condition
\begin{equation}
    X+Y^2+\alpha^2 Z^2=1.\label{cc}
\end{equation}
The critical points of the system (\ref{dynX})-(\ref{dynZ}), their type and dominant contribution in the energy constraint $X+Y^2+\alpha^2 Z^2=1$ are presented in table~\ref{tab:1}.

\begin{figure}[t]
\centering
\includegraphics[scale=0.85]{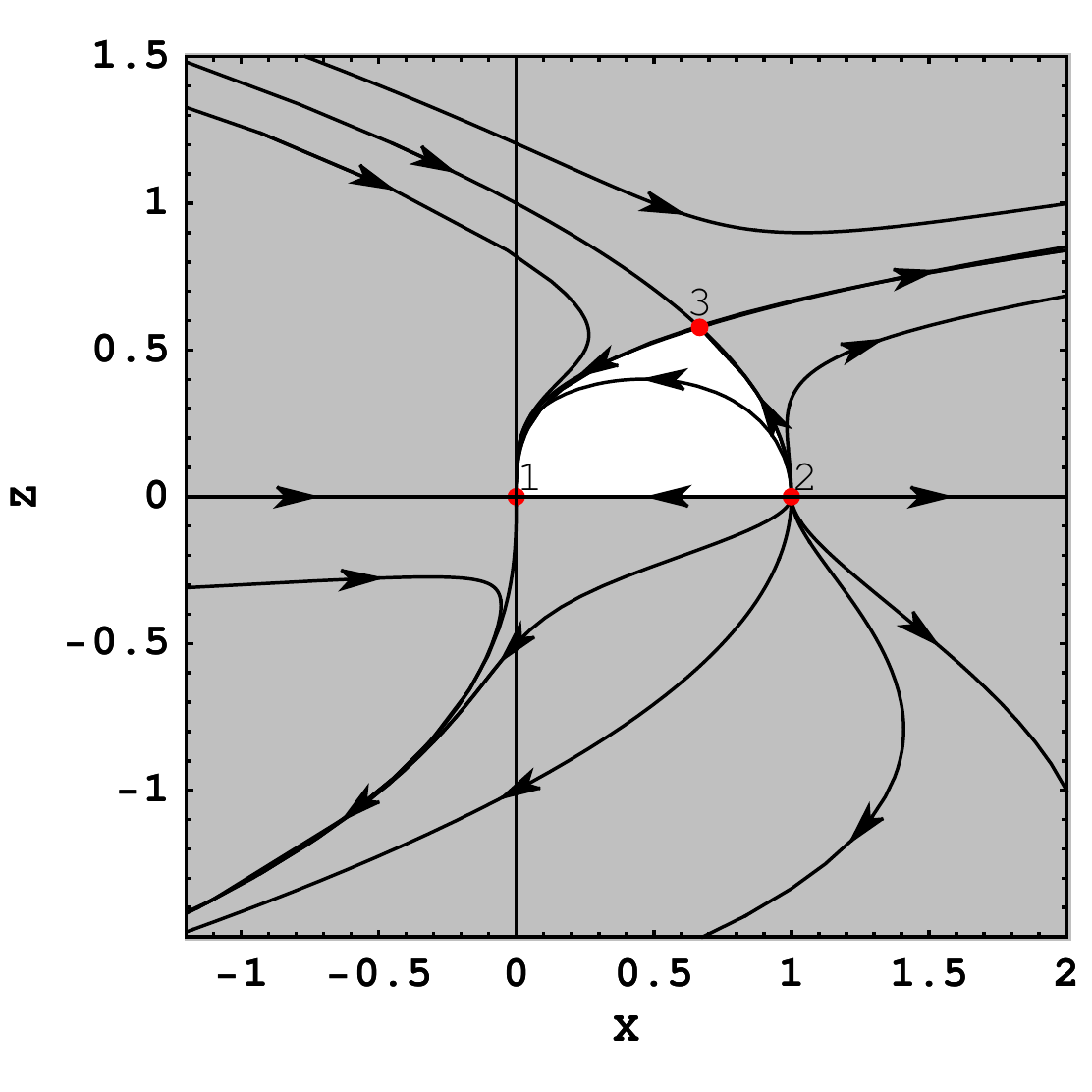}
\caption{The phase portrait for dynamical system (\ref{dynX})-(\ref{dynZ}) for real $\alpha$ and $H>0$. The grey domain represents non-physical solutions. The phase portrait is organized by three critical points: the de Sitter universe represented by a stable node (point 1), the Einstein-de Sitter universe represented by an unstable node (point 2) and the generalization of Einstein-de Sitter represented by a saddle (point 3).}
\label{fig:3}
\end{figure}

\begin{figure}[t]
\centering
\includegraphics[scale=0.85]{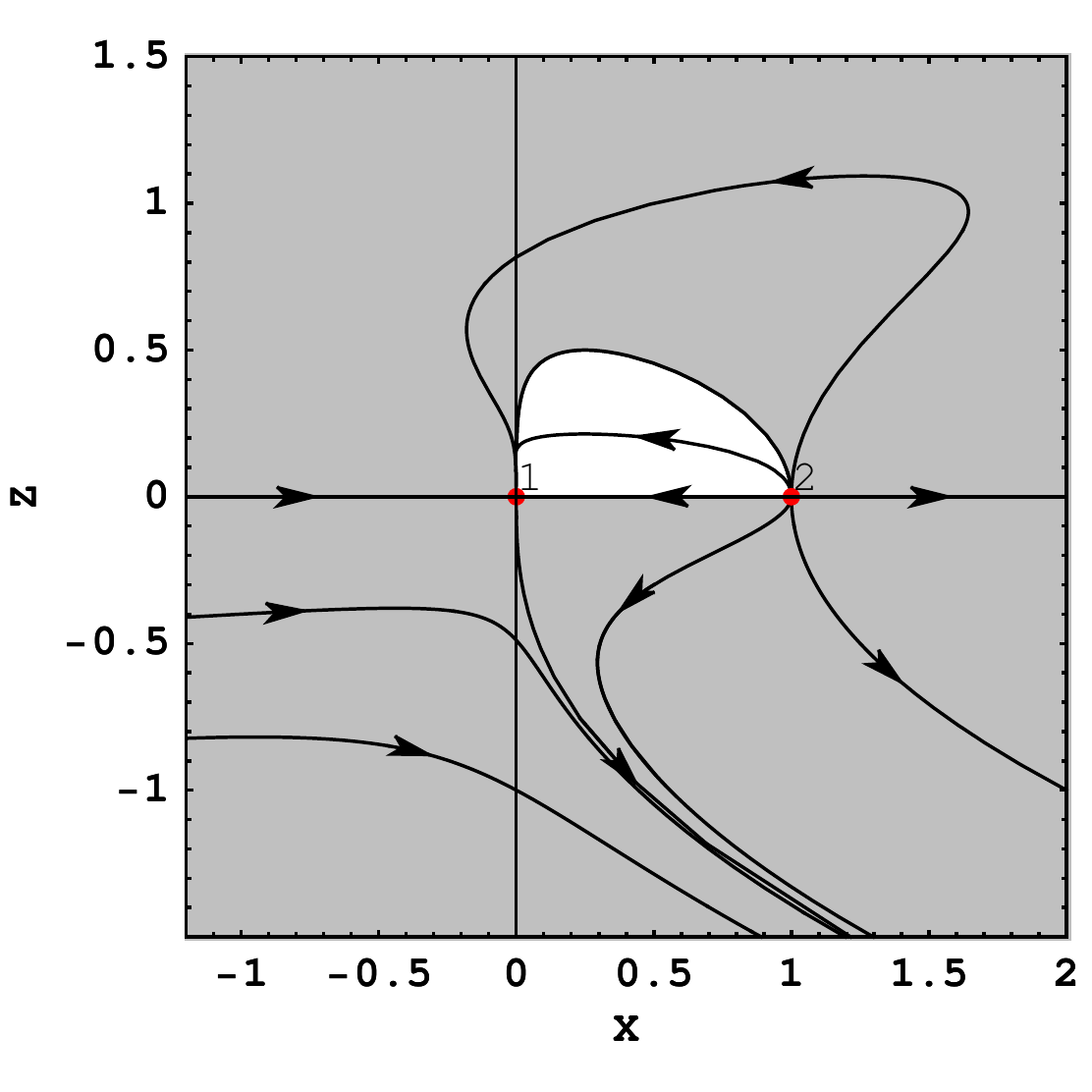}
\caption{The phase portrait for dynamical system (\ref{dynX})-(\ref{dynZ}) for imaginary $\alpha$ and $H>0$. The grey domain represents non-physical solutions. The phase portrait is organized by two critical points: the de Sitter universe represented by a stable node (point 1) and the Einstein-de Sitter universe represented by an unstable node (point 2).}
\label{fig:4}
\end{figure}

System (\ref{dynX})-(\ref{dynZ}) is three-dimensional but it has invariant submanifolds ${Y=0}$ and ${Z=0}$. The behavior of trajectories on the invariant submanifold $Y=0$ describes fully the global dynamic. The phase portraits on the plane ($X$, $Z$) are presented in figures~\ref{fig:3} and \ref{fig:4}. Because of the constraint $Y^2=1-X-\alpha^2 Z^2$ the physical trajectories lie in the region $Y^2\geq 0$. Beyond this region is situated a non-physical region (the shaded region in figures~\ref{fig:3} and \ref{fig:4}). The boundary of the physical region is determined by a parabola $X=1-\alpha^2 Z^2$ and a line $X=0$.

\begin{table}[t]
\caption{The critical points of the system (\ref{dynX})-(\ref{dynZ}), their type and dominant contribution in the energy constraint $X+Y^2+\alpha^2 Z^2=1$.}
\label{tab:1}
\centering
\begin{tabular}{llll}
\hline
No & position of critical point & type & dominant contribution in (\ref{cc})\\[0.1cm] \hline \hline
1 & $X_0=0$, $Y_0=1$, $Z_0=0$ & stable node & $\Lambda$ dominant state in \\
& & & the future (de Sitter) \\[0.2cm] \hline
2 & $X_0=1$, $Y_0=0$, $Z_0=0$ & unstable node & matter dominant state in \\
& & & the past (Einstein-de Sitter) \\[0.2cm] \hline
3 & $X_0=\frac{2}{3\alpha^2}(-1+\sqrt{1+3\alpha^2})$ & saddle & both decaying vacuum effects \\
& $Y_0 = 0$ & & and matter effects \\
& $Z_0=\frac{1}{\sqrt{3}\alpha^2}(-1+\sqrt{1+3\alpha^2})$ & & are dominating in the past \\[0.2cm] \hline
\end{tabular}
\end{table}

\begin{figure}[ht]
\centering
\includegraphics[scale=0.85]{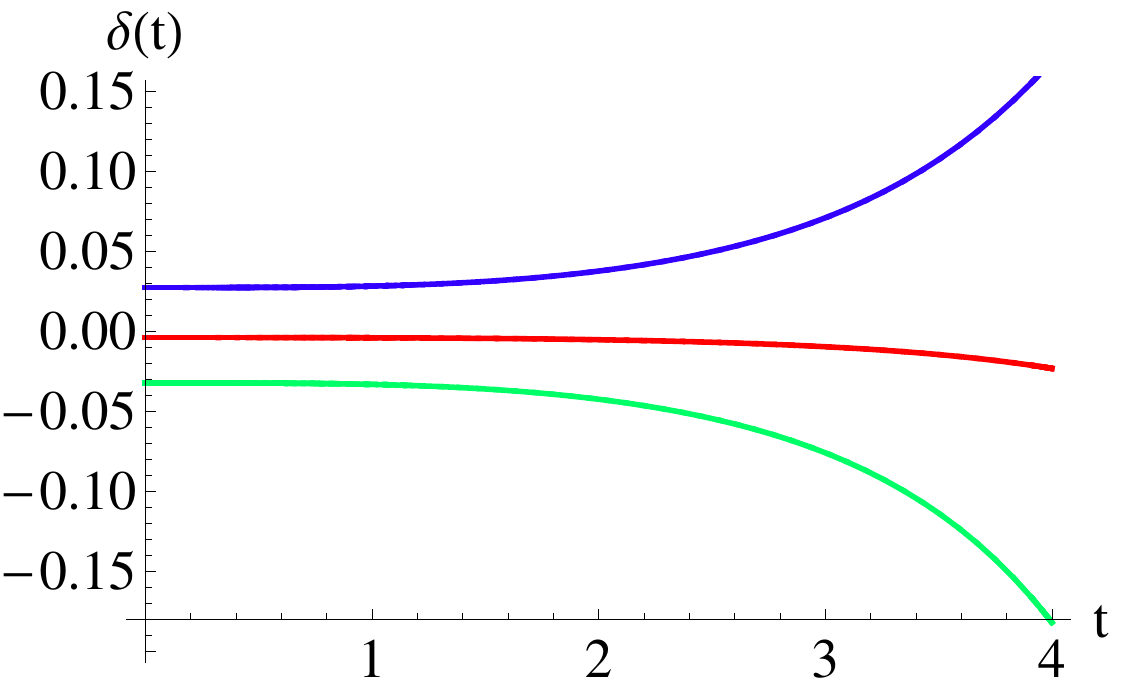}
\caption{Diagram of $\delta(t)$. The top function represents the evolution of $\delta$ assuming the best fit of $H_0=68.27$ km/(s Mpc) and right boundary values of $95\%$ C.L. for $\Omega_{\text{m},0}=0.2416$, $\Omega_\alpha=0.0039$, the middle function represents the best fit for all parameters. The top function represents the evolution of $\delta$ assuming the best of $H_0=68.27$ km/(s Mpc) and the left boundary values of $95\%$ C.L. for $\Omega_{\text{m},0}=0.3542$, $\Omega_{\alpha,0}=-0.0056$. On the $t$-axis we use a unit times $100$ Mpc s/km. See also table~\ref{tab:2}.}
\label{fig:13}
\end{figure}

All critical points lie on this boundary. From the physical point of view they represent asymptotic states of system (\ref{dynX})-(\ref{dynZ}) which are started at $\tau\rightarrow -\infty$ and reach the critical points at $\tau=+\infty$. The critical point marked as (1) represents the de Sitter universe and is a stable node. It is a global attractor for trajectories from its neighborhood. The critical point (2) is an unstable node and it represents the CDM model.

The novelty on the phase portrait is the presence of critical point (3). It is of saddle type. A this critical point $\rho_{\text{m}}= 2\frac{\sqrt{1+3\alpha^2}-1}{\alpha^2} H^2$ and $Ht=\frac{\alpha^2}{\sqrt{1+3\alpha^2}-1}$ i.e., it represents a universe dominated by both decaying vacuum and matter.

The de Sitter state (critical point 1) is connected by an outcoming separatrix with the saddle (point 3). The second separatrix gets in the saddle (point 3) and gets out from the Einstein-de Sitter state (point 2). The other trajectories in a non-shaded region start from the Einstein-de Sitter state and finish in the de Sitter state. In all cases the time flows from $\tau = - \infty$ ($a=0$) to $\tau=+\infty$ ($a=+\infty)$. At the critical point (3) the decaying $\Lambda$ and matter play important role and cannot be neglected.
At this critical point scale factor $a$ and $\rho_{\text{m}}$ behaves like: $a\propto t^{\frac{\alpha^2}{\sqrt{1+3\alpha^2}-1}}$, $H=\frac{\alpha^2}{\sqrt{1+3\alpha^2}-1} t^{-1}$ and $\rho_{\text{m}}=\frac{2\alpha^2}{\sqrt{1+3\alpha^2}-1} t^{-2}$.

This critical point exist only if $\alpha\neq 0$. If $\alpha\rightarrow 0$ then it coincides with the CDM universe. This critical point represents a new generalized CDM model in which $\rho_{\text{m}}=\frac{4}{3} t^{-2}$ and $a(t)\propto t^{\frac{2}{3}}$ in the early universe.

If the function
\begin{multline}
    \delta=-\frac{\frac{d\Lambda(t)}{dt}}{H \rho_\text{m}}=\frac{\frac{2\alpha^2}{t^3}}{H \rho_\text{m}} = -\frac{\alpha^2}{\frac{3}{2}t^3  H_0^3\left(\frac{1-2n}{3H_0 t}+\sqrt{\Omega_{\Lambda,0}}\frac{I_{n-1}\left(\frac{3\sqrt {\Omega_{\Lambda,0}} H_0}{2}t \right)}{
I_n \left(\frac{3\sqrt {\Omega_{\Lambda,0}} H_0}{2}t \right)}\right)} \times \\
\times \left(\Omega_{\Lambda,0}+\frac{\Omega_{\alpha,0}T^2_0}{t^2}-\left(\frac{1-2n}{3H_0 t}+\sqrt{\Omega_{\Lambda,0}}\frac{I_{n-1}\left(\frac{3\sqrt {\Omega_{\Lambda,0}} H_0}{2}t \right)}{
I_n \left(\frac{3\sqrt {\Omega_{\Lambda,0}} H_0}{2}t \right)}\right)^2\right)^{-1}
\end{multline}
is slowly changing then
\begin{equation}
    \rho_{\text{m}}=\rho_{\text{m},0} a^{-3+\delta(t)}.
\end{equation}
Let $\delta(t)=\delta=\text{const}$ then $\rho_{\text{m}}=\rho_{\text{m},0} a^{-3+\delta}$. If $t\rightarrow\infty$ then $\delta(t)\rightarrow \text{const}$.
At the critical point (3) $\delta(t)=\delta=(\sqrt{1+3\alpha^2}-1)^{2}/\alpha^2$. The diagram of $\delta(t)$ is presented in figure~\ref{fig:13}.

Some interesting interpretation of our postulated $\Lambda(t)$ relation can be derive if we apply Starobinsky's argument \cite{Starobinsky:1978SvAL} that $\rho_{\phi}$ after some averaging over time in the interval $\Delta t \gg m^{-1}$ assumes the following form in the quintessence epoch
\begin{equation}
 \rho_{\phi} = V_0 + A a^{-3}.
\end{equation}

Therefore in the matter dominating phase we obtain the $\Lambda(t)$ parametrization (\ref{eq:5}). Finally the model involved belongs to the class of models with so called early dark energy constant in which $\Omega_{\text{de}} = \text{const} \equiv \Omega_e$ during the matter dominated stage (the same refers to the radiation dominated stage, too, but with a different value of $\Omega_e$).

If $\delta \ll 1$ for the fractional density of dark energy $\Omega_\text{e}$ \cite{Doran:2006kp,Pettorino:2013ia} it assumes in the intermediate domain of the universe the following form
\begin{equation}
    1-\Omega_{\text{e}}(a(t))=\frac{\Omega_{\text{m,0}}T_0^2 t^{-2}}{\Omega_{\text{m,0}}T_0^2 t^{-2}+\Omega_{\Lambda,0}+\Omega_{\alpha}T_0^2 t^{-2}}.
\end{equation}
In the early universe this value is constant
\begin{equation}
    \Omega_{e}=\frac{\Omega_{\alpha,0}}{\Omega_{\text{m},0}+\Omega_{\alpha,0}}.
\end{equation}
Therefore for a small value of $\Omega_{\alpha,0}$, $\Omega_{\text{e}}$ is obtained as $\Omega_{\text{e}}=\frac{\Omega_{\alpha,0}}{\Omega_{\text{m}},0}$.

Hojjati et al. \cite{Hojjati:2013oya} found the fraction in total density contributed by early dark energy which is approximately equavalent to $\Omega_e$. 

\begin{figure}[ht]
\centering
\includegraphics[scale=0.85]{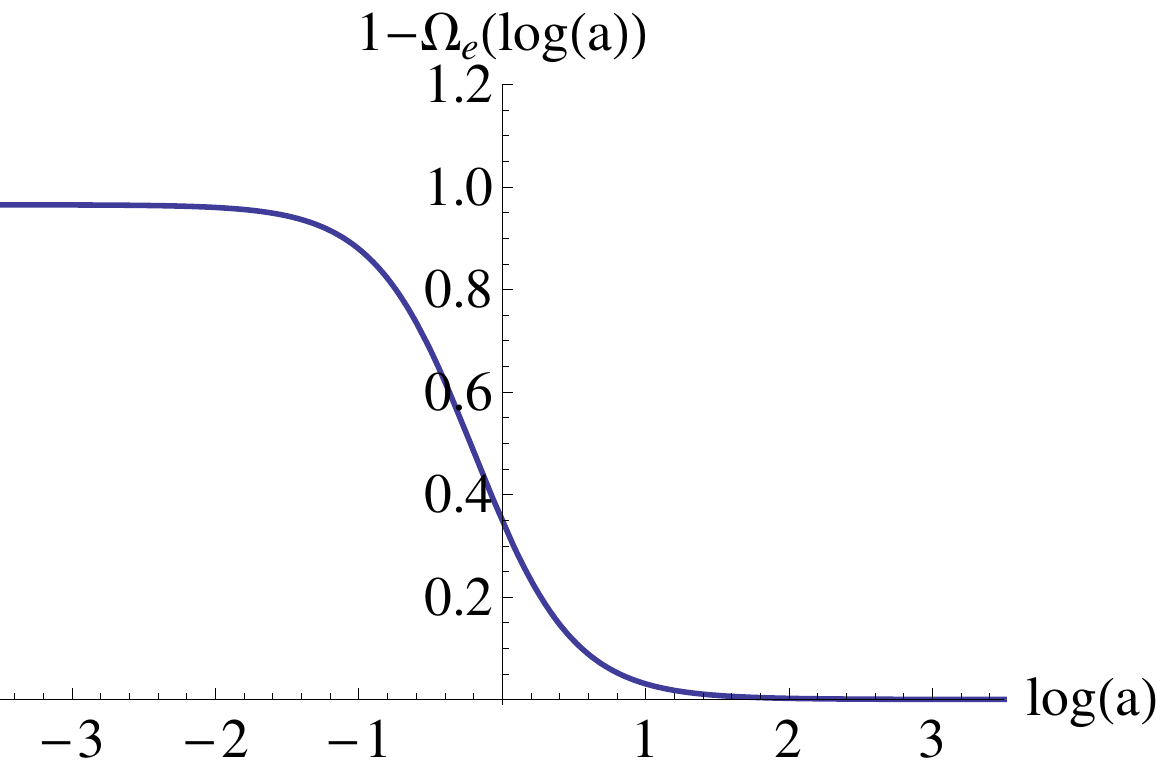}
\caption{Diagram of $1-\Omega_\text{e}(\log(a))$ which is a share of energy density of matter in the total energy density. The function represents the evolution of $1-\Omega_\text{e}$ for $H_0=68.27$ km/(s Mpc), $\Omega_{\text{m},0}=0.35$ and $\Omega_{\alpha,0}=0.05$. The present epoch is at $\log(a) = 0$. For negative values of $\log(a)$ we have past evolution of $1-\Omega_\text{e}(\log(a))$ with a constant phase of the fractional density of dark energy $\Omega_e$ at the early universe ($\Omega_e$ is small and close to zero). For positive values of $\log(a)$ we have future evolution of $1-\Omega_\text{e}(\log(a))$ with a constant phase of the fractional density of dark energy $\Omega_e$ at the late universe ($\Omega_e$ is big and close to one). Between these two constant phases there is an intermediate phase of changing $1-\Omega_\text{e}(\log(a))$ in which we are living.}
\label{fig:15}
\end{figure}

In our model the exact form of $\Omega_{\text{e}}(t)$ is
\begin{equation}
    \Omega_{\text{e}}(t) = \frac{\Omega_{\Lambda,0} + \frac{\Omega_{\alpha,0}T^2_0}{t^2}}{\left(\frac{1-2n}{3H_0 t} + \sqrt{\Omega_{\Lambda,0}}\frac{I_{n-1}\left(\frac{3\sqrt {\Omega_{\Lambda,0}} H_0}{2}t \right)}{I_n \left(\frac{3\sqrt {\Omega_{\Lambda,0}} H_0}{2}t \right)}\right)^2}.
\label{eq:omega_e}
\end{equation}
The evolution of fractional density of dark energy $\Omega_e$ has a shape of the ``logistic'' curve. For convenience, the diagram $1-\Omega_\text{e}(\log(a))$ is presented in figure~\ref{fig:15}.

There is another interesting approach to running cosmologies, proposed by Starobinsky \cite{Starobinsky:1998fr}. In this approach (the bottom up), the universe in the quintessence epoch is described by a scalar field minimally coupled to gravity with some self-interacting potential. He proposed the reconstruction of this potential from the evolution of the scalar perturbation (or also luminosity function) in dust like matter component.

Masso et al. \cite{Masso:2005zg} discussed some aspects of contribution to the dark energy density of coherent scalar field oscillation in the potential. They obtained using the analytical method of adiabatic invariance that for a quadratic potential the energy density $\rho_{\phi}$ evolves as $a^{-3}$ and a quartic potential $V(\phi) \approx \phi^4$ evolves like for the radiation matter $a^{-4}$. Therefore if we add $\Lambda_{\text{bare}}$ to the potential in a matter (or radiation) dominating universe $\rho_{\phi} = \Lambda(t)$.

\section{Statistical analysis of the model}

In this section we present a statistical analysis of the model parameters using the SNIa, BAO, CMB observations, measurements of $H(z)$ and the Alcock-Paczy{\'n}ski test.

First, we use the Union 2.1 sample of 580 supernovae \cite{Suzuki:2011hu}. For the SNIa data we have the following likelihood function
\begin{equation}\label{like_sn}
    \ln L_{\text{SNIa}} = -\frac{1}{2} \sum_{i=1}^{N} \left (\frac{\mu_i^{\text{obs}}-\mu_i^{\text{th}}}{\sigma_i }\right)^2 ,
\end{equation}
where the summing is over the SNIa sample; the distance modulus $\mu^{\text{obs}}=m-M$ (where $m$ is the apparent magnitude and $M$ is the absolute magnitude of SNIa stars) and $\mu^{\text{th}} = 5 \log_{10} D_L +25$ (where the luminosity distance is $D_L= c(1+z) \int_{0}^{z} \frac{d z'}{H(z)}$ and $\sigma$ is the uncertainties.

We use the BAO (baryon acoustic oscillation) data which were taken from the Sloan Digital Sky Survey Release 7 (SDSS DR7) dataset which consists of 893 319 galaxies \cite{Percival:2009xn}. The likelihood function is given by
\begin{equation}
    \ln L_{\text{BAO}} = -\frac{1}{2}\frac{\left( \frac{r_s(z_d)}{D_V(z)} - d(z) \right)^2}{\sigma^2},
\end{equation}
where $r_s(z_d)$ is the sound horizon at the drag epoch and $z=0.275$, $d(z)=0.1390$, $\sigma = 0.0037$ \cite{Eisenstein:1997ik}.

The next likelihood function encompasses the Planck observations of cosmic microwave background (CMB) radiation \cite{Ade:2013zuv}, the information on lensing from the Planck and low-$\ell$ polarization from the WMAP and has the form
\begin{equation}
    \ln L_{\text{CMB}+\text{lensing}+\text{WP}} = - \frac{1}{2} \sum_{ij} (x_i^{\text{th}}-x_i^{\text{obs}}) \mathbb{C}^{-1} (x^{\text{th}}-x^{\text{obs}}),
\end{equation}
where $\mathbb{C}$ is the covariance matrix with the errors, $x$ is a vector of the acoustic scale $l_{A}$, the shift parameter $R$ and $\Omega_{b}h^2$ where
\begin{align}
    l_A &= \frac{\pi}{r_s(z^{*})} c \int_{0}^{z^{*}} \frac{dz'}{H(z')}, \\
    R &= \sqrt{\Omega_{\text{m},0} H_0^2} \int_{0}^{z^{*}} \frac{dz'}{H(z')},
\end{align}
where $z^{*}$ is the recombination redshift and $r_s$ is the sound horizon.

The idea of the Alcock-Paczy{\'n}ski test is the comparison of the radial and tangential size of an object, which is isotropic in the correct choice of model \cite{Alcock:1979mp,Lopez-Corredoira:2013lca}. The likelihood function is independent of the parameter $H_0$ and has the following form
\begin{equation}
    \ln L_{AP} = - \frac{1}{2} \sum_i \frac{\left( AP^{\text{th}}(z_i)-AP^{\text{obs}}(z_i) \right)^2}{\sigma^2},
\end{equation}
where $AP(z)^{\text{th}} \equiv \frac{H(z)}{z} \int_{0}^{z} \frac{dz'}{H(z')}$ and $AP(z_i)^{\text{obs}}$ are observational data \cite{Sutter:2012tf,Blake:2011ep,Ross:2006me,Marinoni:2010yoa,daAngela:2005gk,Outram:2003ew,Anderson:2012sa,Paris:2012iw,Schneider:2010hm}.

At the end it is also valuable to add the constraints on the Hubble parameter, i.e. $H(z=0)\equiv H_0$.

Data of $H(z)$ for samples of different galaxies were also used \cite{Simon:2004tf,Stern:2009ep,Moresco:2012jh}.
\begin{equation}\label{hz}
    \ln L_{H(z)} = -\frac{1}{2} \sum_{i=1}^{N} \left (\frac{H(z_i)^{\text{obs}}-H(z_i)^{\text{th}}}{\sigma_i }\right)^2.
\end{equation}

The final likelihood function for the observational Hubble function is
\begin{equation}
    L_{\text{tot}} = L_{\text{SNIa}} L_{\text{BAO}} L_{\text{CMB}+\text{lensing}+\text{WP}} L_{\text{AP}} L_{H(z)}.
\end{equation}

\begin{figure}[ht]
\centering
\includegraphics[scale=0.85]{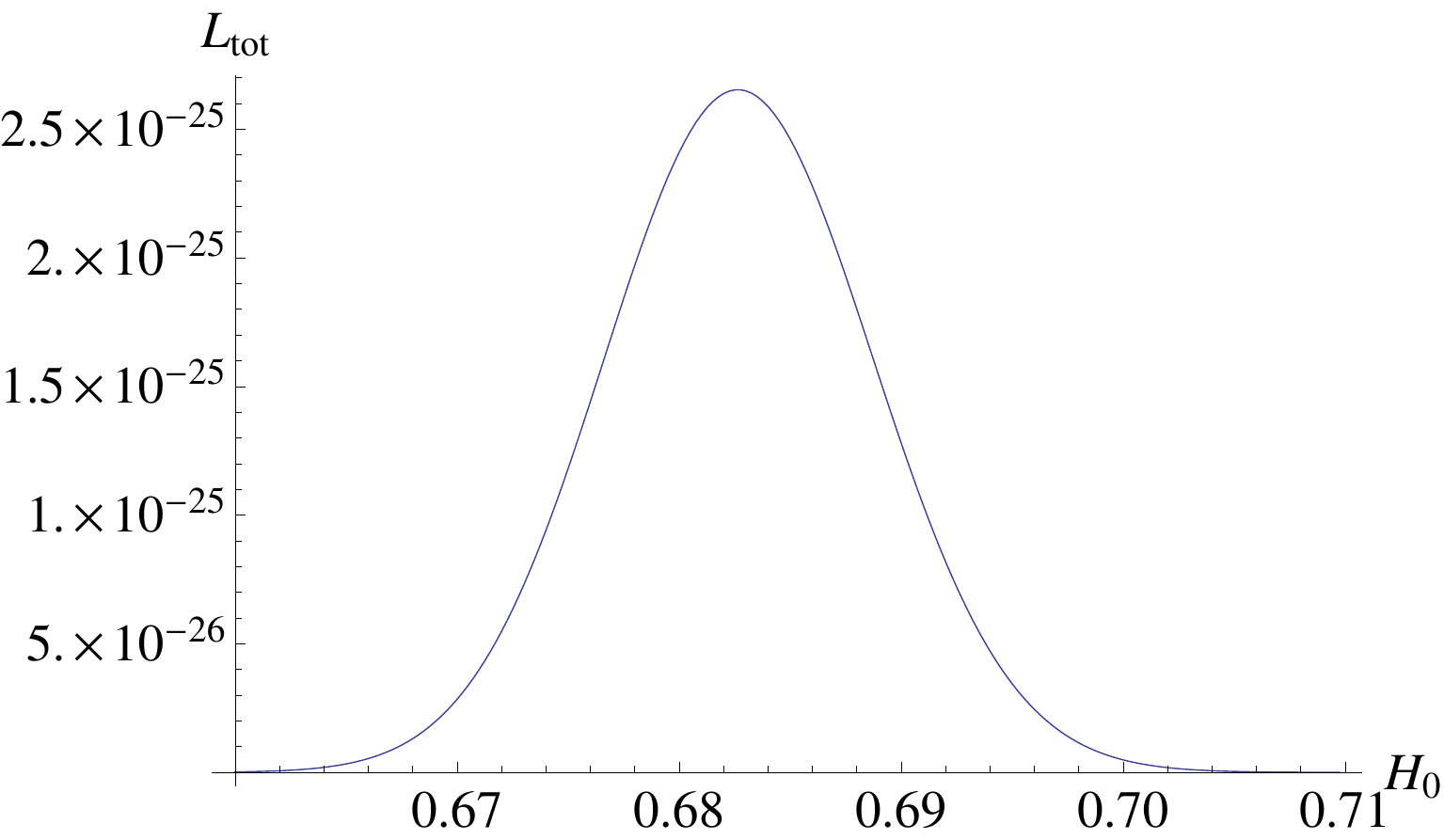}
\caption{Diagram of PDF for parameter $H_0$ in units km/(100 s Mpc) obtained as an intersection of a likelihood function. Two planes of intersection likelihood function are $\Omega_{\text{m},0}=0.2938$ and $\Omega_{\alpha,0}=-0.0006$.}
\label{fig:7}
\end{figure}
\begin{figure}[ht]
\centering
\includegraphics[scale=0.85]{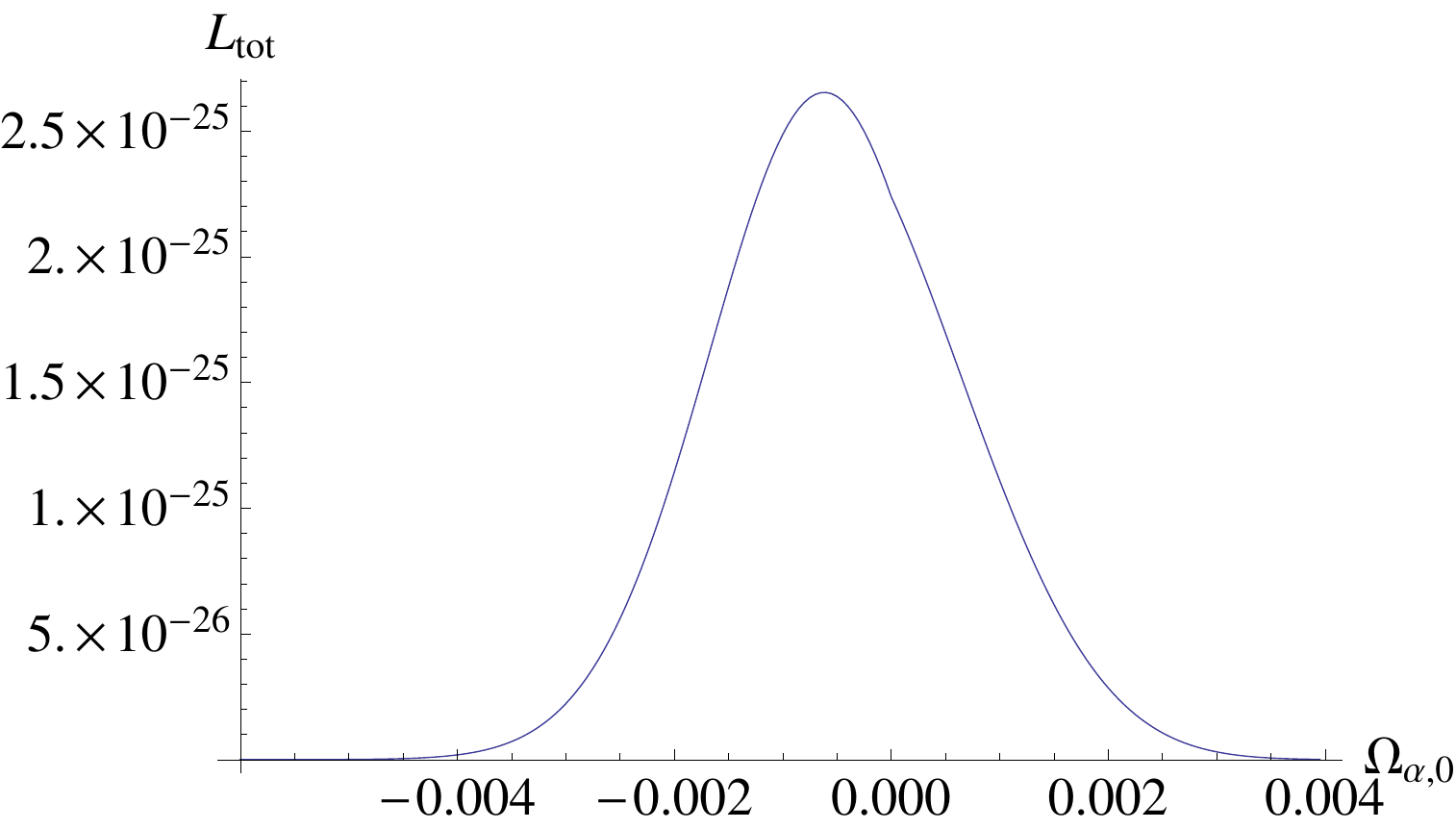}
\caption{Diagram of PDF for parameter $\Omega_{\alpha,0}$ obtained as an intersection of a likelihood function. Two planes of intersection likelihood function are $\Omega_{\text{m},0}=0.2938$ and $H_0=68.27$ km/(s Mpc).}
\label{fig:8}
\end{figure}

\begin{figure}[ht]
\centering
\includegraphics[scale=0.85]{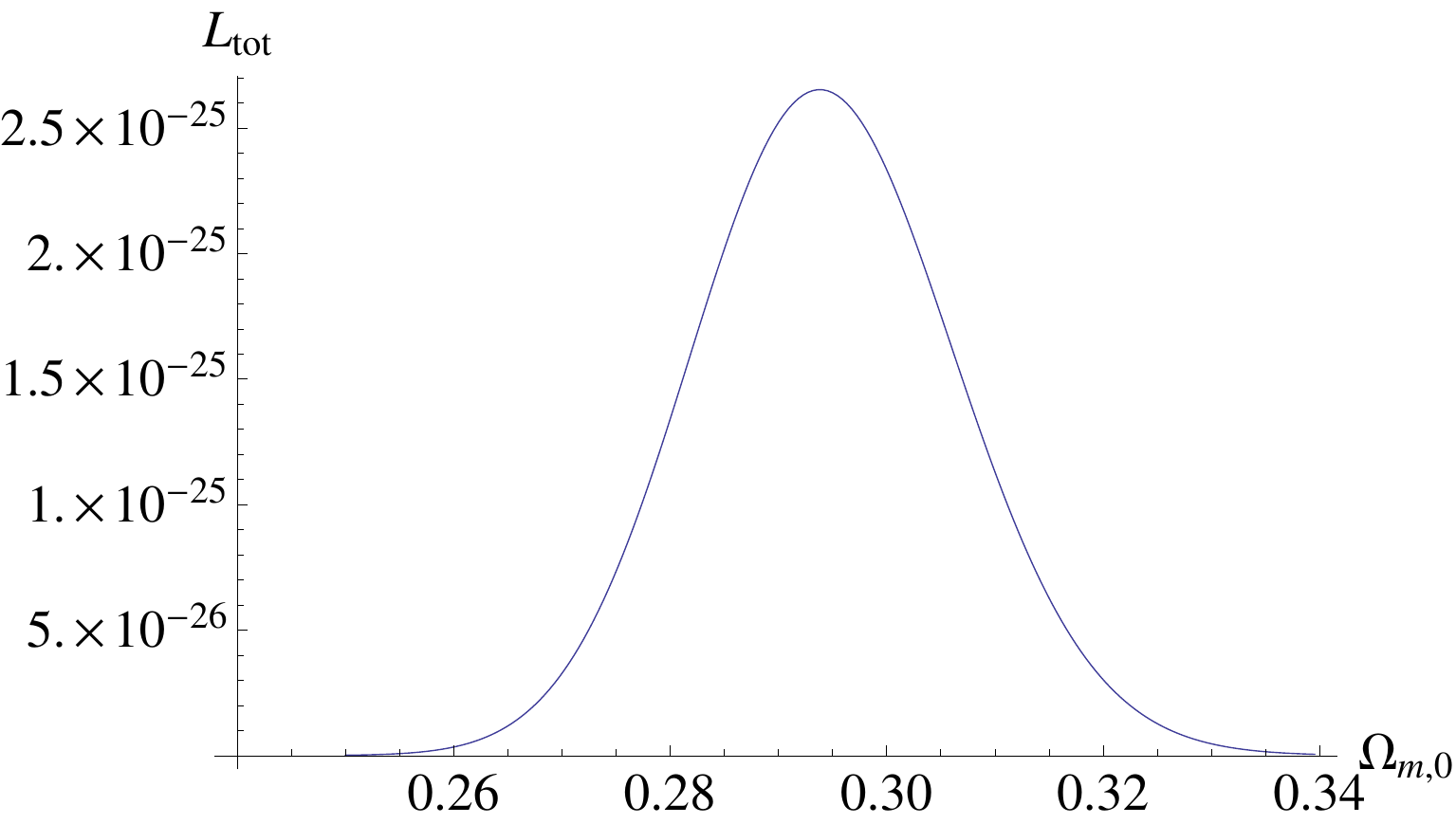}
\caption{Diagram of PDF for parameter $\Omega_{\text{m},0}$ obtained as an intersection of a likelihood function. Two planes of intersection likelihood function are $\Omega_{\alpha,0}=-0.0006$ and $H_0=68.27$ km/(s Mpc).}
\label{fig:9}
\end{figure}

\begin{figure}[ht]
\centering
\includegraphics[scale=0.85]{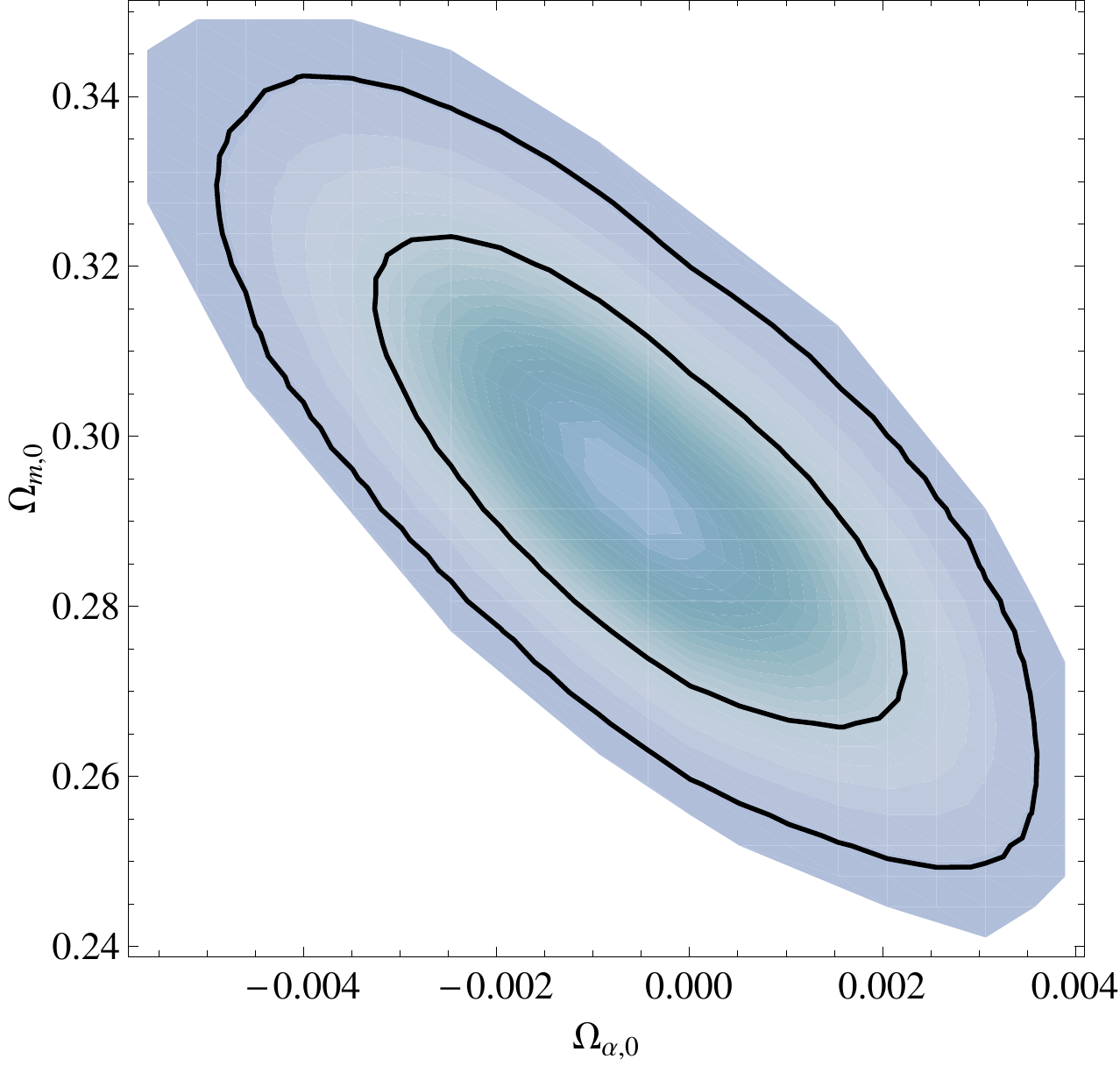}
\caption{The likelihood function of two model parameters ($\Omega_{\alpha,0}, \Omega_{\text{m},0}$) with the marked $68\%$ and $95\%$ confidence levels. The value of Hubble constant is estimated from the data as best fit value $H_0 = 68.27$ km/(s Mpc) and then the diagram of likelihood function is obtained for this value.}
\label{fig:5}
\end{figure}

\begin{figure}[ht]
\centering
\includegraphics[scale=0.85]{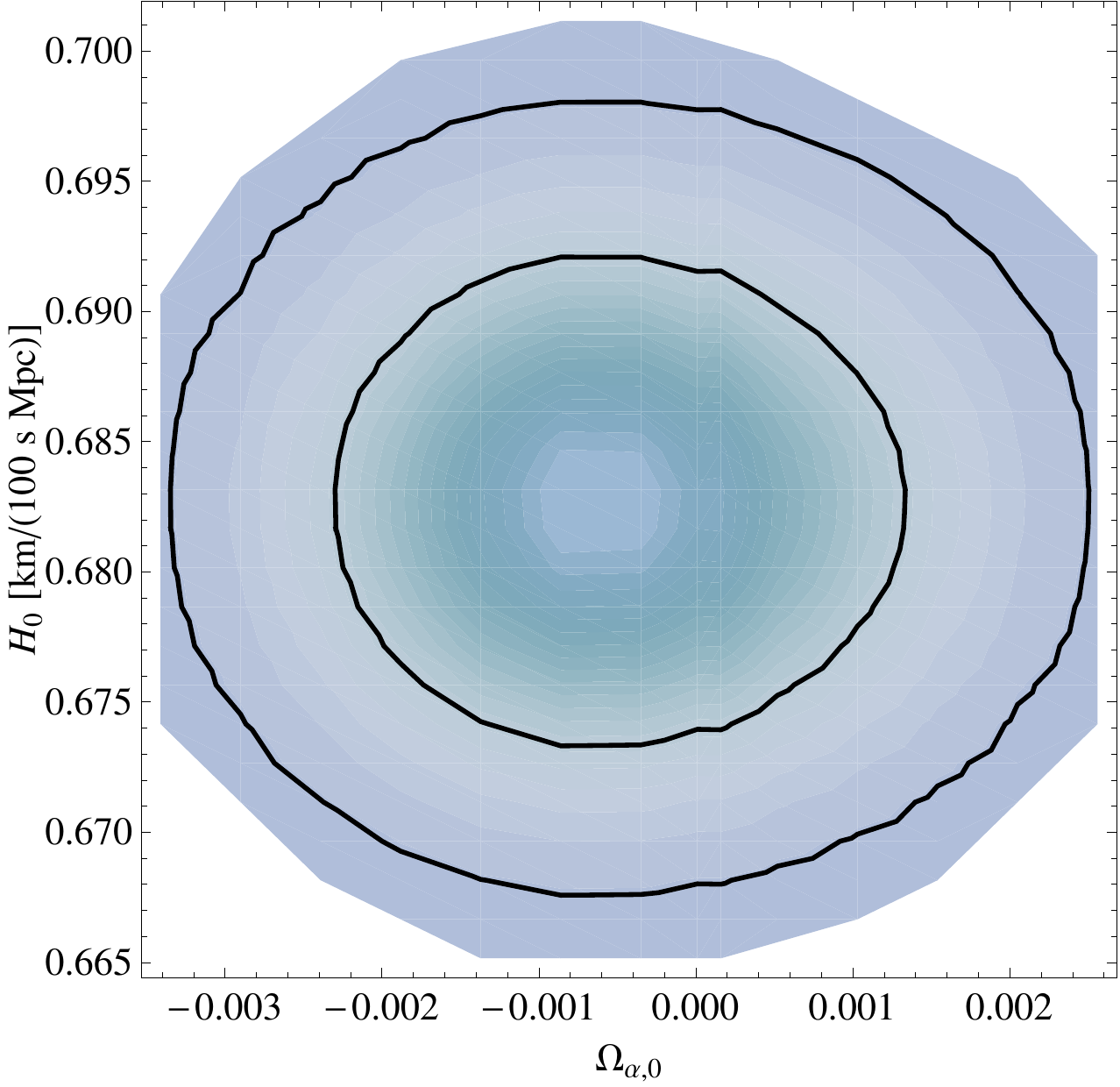}
\caption{The likelihood function of two model parameters ($\Omega_{\alpha,0}, H_{0}$) with the marked $68\%$ and $95\%$ confidence levels. The value of $\Omega_{\text{m},0}$ constant is estimated from the data as best fit value $\Omega_{\text{m},0} = 0.2938$ and then the diagram of likelihood function is obtained for this value.}
\label{fig:6}
\end{figure}

\begin{table}
\caption{The best fit and errors for the estimated model with $\alpha^2$ from the interval $(-0.05, 0.05)$.}
\label{tab:2}
\begin{center}
\begin{tabular}{llll} \hline
parameter & best fit & $ 68\% $ CL & $ 95\% $ CL \\ \hline \hline
$H_0$ & 68.27 km/(s Mpc) & $\begin{array}{c}
  +0.96 \\ -1.07
\end{array}$ & $\begin{array}{c}
  +1.67 \\ -1.68
\end{array}$ \\ \hline
$\Omega_{m,0}$ & 0.2938 & $\begin{array}{c}
  +0.0355 \\ -0.0325
\end{array}$ & $\begin{array}{c}
  +0.0604 \\ -0.0522
\end{array}$\\ \hline
$\Omega_{\alpha,0}$ & -0.0006 & $\begin{array}{c}
  +0.0031 \\ -0.0030
\end{array}$ & $\begin{array}{c}
  +0.0045 \\ -0.0050
\end{array}$ \\ \hline
$j_0$ & 1.002 & $\begin{array}{c}
  +0.010 \\ -0.008
\end{array}$ & $\begin{array}{c}
  +0.016 \\ -0.013
\end{array}$ \\ \hline
$\alpha^2$ & -0.002 & $\begin{array}{c}
  +0.010 \\ -0.007
\end{array}$ & $\begin{array}{c}
  +0.014 \\ -0.012
\end{array}$ \\ \hline
\end{tabular}
\end{center}
\end{table}

\begin{table}
\caption{The best fit and errors for the estimated model with positive $\alpha^2$ from the interval $(0.00, 0.05)$.}
\label{tab:3}
\begin{center}
\begin{tabular}{llll} \hline
parameter & best fit & $ 68\% $ CL & $ 95\% $ CL \\ \hline \hline
$H_0$ & $68.38$ km/(s Mpc)& $\begin{array}{c}
  +0.97 \\ -0.97
\end{array}$ & $\begin{array}{c}
  +1.16 \\ -1.16
\end{array}$ \\ \hline
$\Omega_{\text{m},0}$ & 0.2877 & $\begin{array}{c}
  +0.0198 \\ -0.0242
\end{array}$ & $\begin{array}{c}
  +0.0331 \\ -0.0442
\end{array}$\\ \hline
$\Omega_{\alpha,0}$ & 0.0000 & $\begin{array}{c}
  +0.0025 \\ -0.0000
\end{array}$ & $\begin{array}{c}
  +0.0038 \\ -0.0000
\end{array}$ \\ \hline
$j_0$ & 1.000 & $\begin{array}{c}
  +0.000 \\ -0.008
\end{array}$ & $\begin{array}{c}
  +0.000 \\ -0.011
\end{array}$ \\ \hline
$\alpha^2$ & 0.000 & $\begin{array}{c}
  +0.008 \\ -0.000
\end{array}$ & $\begin{array}{c}
  +0.012 \\ -0.000
\end{array}$ \\ \hline
\end{tabular}
\end{center}
\end{table}

To estimate the model parameters we use our own code CosmoDarkBox implementing the Metropolis-Hastings algorithm \cite{Hastings:1970aa}, \cite{Metropolis:1953am}. We use the dynamical system formulation of model to obtain the likelihood function \cite{Hu:1995en,Eisenstein:1997ik}.

We use observation data of 580 supernovae of type Ia, selected subsets of the data points of Hubble function, the measurements of BAO from SDSS DR7. We also use data for the application of the Alcock-Paczy{\'n}ski test 18 observational points. At last, we estimated model parameters with CMB data from Planck, low- $\ell$ polarization from WMAP and lensing from Planck. To estimate the model parameters we chose interval (64.00, 74.00) for $H_0$ and (0.21, 0.37) for $\Omega_{\text{m},0}$. The values of estimated parameters for $\alpha^2$ from the interval (-0.05, 0.05) are shown in table~\ref{tab:2} and for positive $\alpha^2$ from the interval (0.00, 0.05) are shown in table~\ref{tab:3}. The best fit for model with $\alpha^2$ from the interval (-0.05, 0.05) is in the part of likelihood function where $\alpha^2$ is negative.

If it is chosen the lower limit of the interval of $\alpha^2$ larger than the value of the best fit then the best fit of the model for the new interval is equal the value of the lower limit of this interval. A consequence is the lower limit of the error is equal zero for $\alpha^2$. So the specific values of the best fit and errors of $\Omega_{\alpha,0}$, $\alpha^2$ and $j_0$ in table~\ref{tab:3} are a result of the choice of limits of the interval of $\alpha^2$.

To illustrate the results of statistical analysis the diagrams of PDF are shown in figures~\ref{fig:7}, \ref{fig:8} and~\ref{fig:9}. In turn figures~\ref{fig:5} and \ref{fig:6} shown the likelihood function with $68\%$ and $95\%$ confidence level projection on the ($\Omega_{\alpha,0}$, $\Omega_{\text{m},0}$) plane and the ($\Omega_{\alpha,0}$, $H_0$) plane, respectively.

\section{Conclusion}

The aim of the paper was to study the dynamics of the emerging $\Lambda(t)$CDM cosmological models. In the study of dynamics we find exact solutions and use dynamical system methods for the analysis of dependence of solutions on initial conditions. In the latter evolutional paths of cosmological model are represented by trajectories in the phase space. Due to geometrical visualization of dynamics we have the space of all solutions and can discuss their stability. We are looking for such trajectories for which the $\Lambda$CDM model is a global attractor in the phase space.

We study in details dynamics of cosmological model with the prescribed form of $\Lambda(t)=\Lambda_{\text{bare}}+\frac{\alpha^2}{t^2}$, where $\Lambda_{\text{bare}}$ is a positive constant and $\alpha^2$ is either positive or negative. We calculate exact solutions for the scale factor and subsequently calculate the jerk. It is demonstrated that this parameter is time dependent if and only if the effects of time contribution to $\Lambda(t)$ are non-zero. We propose the measurement of the jerk as a diagnostic of decaying $\Lambda$, i.e. $\dot\Lambda<0$. Due to analysis of dynamics in the phase space we have found an interesting solution in the phase space, a saddle critical point, at which $\rho_{\text{m}}(t)$ scales like $t^{-2}$. This solution was recently proposed by Ringermacher and Mead \cite{Ringermacher:2014kaa} as a description characteristic for the dark matter evolution.

From the phase portrait we derive the generic scenario for an evolution of cosmological models with such a form of the dark energy parametrization. Trajectories start from initial singularity (the Einstein-de Sitter model) and then go in vicinity of the saddle point where they spend a lot of time and then go to the de Sitter state. It is a typical behavior for all generic trajectories in the phase space. The new critical point is emerging in the phase space due to the effect of a time dependence of the cosmological constant. If $\alpha^2 \leq 0$ then this point is absent (it is gluing with the critical point representing the Einstein-de Sitter model).

We also tested this model using astronomical data. Statistical estimations show that the model fits to data as well as the standard cosmological model (the $\Lambda$CDM model). In any case the value of $\alpha^2=0$ belongs to the confidence interval for the estimated parameter $\alpha^2$ we cannot reject that $\alpha^2 \neq 0$. Only if we find the best fit value $\alpha^2$ with the error of the one order less than this value the problem of $\alpha^{2}_{\text{estimated}} \neq \alpha_{0}^{2} \neq 0$ could be solvable. We can obtain the limits on the value of parameter $\alpha^2$ and $-0.009<\alpha^2=3H_0^2T_0^2(1-\Omega_{\text{m},0}-\Omega_{\Lambda,0})<0.008$ (for 68\% C.L.) and $-0.014<\alpha^2 < 0.012$ (for 95\% C.L.).

In papers of Doran and Robbers \cite{Doran:2006kp} and Pettorino et al. \cite{Pettorino:2013ia} there are limits on fractional dark energy at early time. Recently Ade et al. \cite{Ade:2015rim} have found $\Omega_{e} < 0.0036$. It is interesting that they have obtained a similar limit to our limit on $\Omega_{\alpha} < 0.0038$ in the other parametrization of dark energy.

Note that if we apply Starobinsky's idea the parameter $\alpha^2$ can be constrained through $\Omega_{\text{e}}$ measurement. This parameter measures amount of dark energy at the early evolution of the Universe. If $\Omega_{\text{e}}$ is different from zero then we obtain value information about this alternative evolutional scenarios which are consistent with the present epoch.

In our case $\Omega_{\text{e}}=\frac{\Omega_{\alpha,0}}{\Omega_{\text{m,0}}}=\frac{\frac{\alpha^2}{3H_0^2 T_0^2}}{\Omega_{m,0}}<0.0036$ and therefore $\alpha^2<3 H_0^2 T_0^2 \Omega_{\text{m},0}\Omega_{\text{e}}$. If we put $\Omega_{\text{m},0}=0.25$ and $H_0^2 T_0^2=1$ then we obtain $\alpha^2<\frac{3}{4}\Omega_{e}=0.0027$.

Finally we obtain a stronger limit for $\alpha^2$ then in table~\ref{tab:3}. However, note that this estimation is model dependent (it is assumed Starobinsky's argument). note that the case study of our model fully confirm existence of phase during the early universe at which fractional energy density of dark energy is constant (see figure~\ref{fig:15} and eq.~\ref{eq:omega_e}).

\subsection*{Acknowledgements}
The work was supported by the grant NCN DEC-2013/09/B/ST2/03455. We are very grateful of prof. A. Borowiec, Z. Haba, A. Krawiec and K. Urbanowski for stimulating discussion and remarks. Especially I would like to thank S. Odintsov and V. Oikonomou for discussion of the problem of a covariance of the vacuum. We also thank referees for their remarks, especially for indicating the possibility of estimation $\alpha^2$ from measurement of the fractional dark energy density $\Omega_e$.


\begin{thebibliography}{10}

\bibitem{Pani:2013qfa}
P.~Pani, T.~P. Sotiriou, and D.~Vernieri, {\it {Gravity with Auxiliary
  Fields}},  {\em Phys.Rev.} {\bf D88} (2013), no.~12 121502,
  [\href{http://arxiv.org/abs/1306.1835}{{\tt arXiv:1306.1835}}].

\bibitem{Szydlowski:2015bwa}
M.~Szydlowski, {\it {Cosmological model with decaying vacuum energy from
  quantum mechanics}},  {\em Phys. Rev.} {\bf D91} (2015), no.~12 123538,
  [\href{http://arxiv.org/abs/1502.04737}{{\tt arXiv:1502.04737}}].

\bibitem{Lima:1995kd}
J.~Lima, {\it {Thermodynamics of decaying vacuum cosmologies}},  {\em
  Phys.Rev.} {\bf D54} (1996) 2571--2577,
  [\href{http://arxiv.org/abs/gr-qc/9605055}{{\tt gr-qc/9605055}}].

\bibitem{Lima:2015kda}
J.~Lima, E.~Perico, and G.~Zilioti, {\it {Decaying Vacuum Inflationary
  Cosmologies: A Complete Scenario Including Curvature Effects}},
  \href{http://arxiv.org/abs/1502.01913}{{\tt arXiv:1502.01913}}.

\bibitem{Perico:2013mna}
E.~Perico, J.~Lima, S.~Basilakos, and J.~Sola, {\it {Complete Cosmic History
  with a dynamical $\Lambda=\Lambda(H)$ term}},  {\em Phys.Rev.} {\bf D88}
  (2013), no.~6 063531, [\href{http://arxiv.org/abs/1306.0591}{{\tt
  arXiv:1306.0591}}].

\bibitem{Szydlowski:2015kqa}
M.~Szydlowski, A.~Stachowski, and K.~Urbanowski, {\it {Cosmology with a
  Decaying Vacuum Energy Parametrization Derived from Quantum Mechanics}},
  {\em J. Phys. Conf. Ser.} {\bf 626} (2015), no.~1 012033,
  [\href{http://arxiv.org/abs/1502.04471}{{\tt arXiv:1502.04471}}].

\bibitem{Urbanowski:2013tfa}
K.~Urbanowski and K.~Raczynska, {\it {Possible Emission of Cosmic $X$- and
  $\gamma$-rays by Unstable Particles at Late Times}},  {\em Phys. Lett.} {\bf
  B731} (2014) 236--241, [\href{http://arxiv.org/abs/1303.6975}{{\tt
  arXiv:1303.6975}}].

\bibitem{Urbanowski:2012pka}
K.~Urbanowski and M.~Szydlowski, {\it {Cosmology with a decaying vacuum}},
  {\em AIP Conf.Proc.} {\bf 1514} (2012) 143--146,
  [\href{http://arxiv.org/abs/1304.2796}{{\tt arXiv:1304.2796}}].

\bibitem{Cai:2007us}
R.-G. Cai, {\it {A Dark Energy Model Characterized by the Age of the
  Universe}},  {\em Phys.Lett.} {\bf B657} (2007) 228--231,
  [\href{http://arxiv.org/abs/0707.4049}{{\tt arXiv:0707.4049}}].

\bibitem{Wei:2007ty}
H.~Wei and R.-G. Cai, {\it {A New Model of Agegraphic Dark Energy}},  {\em
  Phys.Lett.} {\bf B660} (2008) 113--117,
  [\href{http://arxiv.org/abs/0708.0884}{{\tt arXiv:0708.0884}}].

\bibitem{Maziashvili:2007dk}
M.~Maziashvili, {\it {Cosmological implications of Karolyhazy uncertainty
  relation}},  {\em Phys. Lett.} {\bf B652} (2007) 165--168,
  [\href{http://arxiv.org/abs/0705.0924}{{\tt arXiv:0705.0924}}].

\bibitem{Chen:2011rz}
Y.~Chen, Z.-H. Zhu, L.~Xu, and J.~Alcaniz, {\it {$\Lambda(t)$CDM Model as a
  Unified Origin of Holographic and Agegraphic Dark Energy Models}},  {\em
  Phys.Lett.} {\bf B698} (2011) 175--182,
  [\href{http://arxiv.org/abs/1103.2512}{{\tt arXiv:1103.2512}}].

\bibitem{Li:2004rb}
M.~Li, {\it {A Model of holographic dark energy}},  {\em Phys.Lett.} {\bf B603}
  (2004) 1, [\href{http://arxiv.org/abs/hep-th/0403127}{{\tt hep-th/0403127}}].

\bibitem{Zhang:2012pr}
J.-F. Zhang, Y.-H. Li, and X.~Zhang, {\it {A global fit study on the new
  agegraphic dark energy model}},  {\em Eur.Phys.J.} {\bf C73} (2013), no.~1
  2280, [\href{http://arxiv.org/abs/1212.0300}{{\tt arXiv:1212.0300}}].

\bibitem{Boehmer:2015kta}
C.~G. Boehmer, N.~Tamanini, and M.~Wright, {\it {Interacting quintessence from
  a variational approach Part I: algebraic couplings}},  {\em Phys. Rev.} {\bf
  D91} (2015), no.~12 123002, [\href{http://arxiv.org/abs/1501.06540}{{\tt
  arXiv:1501.06540}}].

\bibitem{Ringermacher:2014kaa}
H.~I. Ringermacher and L.~R. Mead, {\it {Model-Independent Plotting of the
  Cosmological Scale Factor as a Function of Lookback Time}},  {\em Astron.J.}
  {\bf 148} (2014) 94, [\href{http://arxiv.org/abs/1407.6300}{{\tt
  arXiv:1407.6300}}].

\bibitem{Haba:2013xka}
Z.~Haba, {\it {Einstein gravity of a diffusing fluid}},  {\em
  Class.Quant.Grav.} {\bf 31} (2014) 075011,
  [\href{http://arxiv.org/abs/1307.8150}{{\tt arXiv:1307.8150}}].

\bibitem{Haba:2014fd}
Z.~Haba, {\it A fluid of diffusing particles and its cosmological behaviour},
  in {\em Proc. of Sci.: Frontiers of Fundamental Physics 14}.
\newblock 2014.
\newblock ArticleID 193, Conference held 15-18 July 2014, Marseille, France.

\bibitem{Sahni:2008xx}
V.~Sahni, A.~Shafieloo, and A.~A. Starobinsky, {\it {Two new diagnostics of
  dark energy}},  {\em Phys.Rev.} {\bf D78} (2008) 103502,
  [\href{http://arxiv.org/abs/0807.3548}{{\tt arXiv:0807.3548}}].

\bibitem{Shafieloo:2012rs}
A.~Shafieloo, V.~Sahni, and A.~A. Starobinsky, {\it {A new null diagnostic
  customized for reconstructing the properties of dark energy from BAO data}},
  {\em Phys. Rev.} {\bf D86} (2012) 103527,
  [\href{http://arxiv.org/abs/1205.2870}{{\tt arXiv:1205.2870}}].

\bibitem{Sahni:2014ooa}
V.~Sahni, A.~Shafieloo, and A.~A. Starobinsky, {\it {Model independent evidence
  for dark energy evolution from Baryon Acoustic Oscillations}},  {\em
  Astrophys. J.} {\bf 793} (2014), no.~2 L40,
  [\href{http://arxiv.org/abs/1406.2209}{{\tt arXiv:1406.2209}}].

\bibitem{Starobinsky:1978SvAL}
A.~A. Starobinskii, {\it {On a nonsingular isotropic cosmological model}},
  {\em Soviet Astronomy Letters} {\bf 4} (Feb., 1978) 82--84.

\bibitem{Doran:2006kp}
M.~Doran and G.~Robbers, {\it {Early dark energy cosmologies}},  {\em JCAP}
  {\bf 0606} (2006) 026, [\href{http://arxiv.org/abs/astro-ph/0601544}{{\tt
  astro-ph/0601544}}].

\bibitem{Pettorino:2013ia}
V.~Pettorino, L.~Amendola, and C.~Wetterich, {\it {How early is early dark
  energy?}},  {\em Phys. Rev.} {\bf D87} (2013) 083009,
  [\href{http://arxiv.org/abs/1301.5279}{{\tt arXiv:1301.5279}}].

\bibitem{Hojjati:2013oya}
A.~Hojjati, E.~V. Linder, and J.~Samsing, {\it {New Constraints on the Early
  Expansion History of the Universe}},  {\em Phys. Rev. Lett.} {\bf 111}
  (2013), no.~4 041301, [\href{http://arxiv.org/abs/1304.3724}{{\tt
  arXiv:1304.3724}}].

\bibitem{Starobinsky:1998fr}
A.~A. Starobinsky, {\it {How to determine an effective potential for a variable
  cosmological term}},  {\em JETP Lett.} {\bf 68} (1998) 757--763,
  [\href{http://arxiv.org/abs/astro-ph/9810431}{{\tt astro-ph/9810431}}].
  [Pisma Zh. Eksp. Teor. Fiz.68,721(1998)].

\bibitem{Masso:2005zg}
E.~Masso, F.~Rota, and G.~Zsembinszki, {\it {Scalar field oscillations
  contributing to dark energy}},  {\em Phys. Rev.} {\bf D72} (2005) 084007,
  [\href{http://arxiv.org/abs/astro-ph/0501381}{{\tt astro-ph/0501381}}].

\bibitem{Suzuki:2011hu}
N.~Suzuki et~al., {\it {The Hubble Space Telescope Cluster Supernova Survey: V.
  Improving the Dark Energy Constraints Above z>1 and Building an
  Early-Type-Hosted Supernova Sample}},  {\em Astrophys. J.} {\bf 746} (2012)
  85, [\href{http://arxiv.org/abs/1105.3470}{{\tt arXiv:1105.3470}}].

\bibitem{Percival:2009xn}
{\bf SDSS} Collaboration, W.~J. Percival et~al., {\it {Baryon Acoustic
  Oscillations in the Sloan Digital Sky Survey Data Release 7 Galaxy Sample}},
  {\em Mon. Not. Roy. Astron. Soc.} {\bf 401} (2010) 2148--2168,
  [\href{http://arxiv.org/abs/0907.1660}{{\tt arXiv:0907.1660}}].

\bibitem{Eisenstein:1997ik}
D.~J. Eisenstein and W.~Hu, {\it {Baryonic features in the matter transfer
  function}},  {\em Astrophys. J.} {\bf 496} (1998) 605,
  [\href{http://arxiv.org/abs/astro-ph/9709112}{{\tt astro-ph/9709112}}].

\bibitem{Ade:2013zuv}
{\bf Planck} Collaboration, P.~A.~R. Ade et~al., {\it {Planck 2013 results.
  XVI. Cosmological parameters}},  {\em Astron. Astrophys.} {\bf 571} (2014)
  A16, [\href{http://arxiv.org/abs/1303.5076}{{\tt arXiv:1303.5076}}].

\bibitem{Alcock:1979mp}
C.~Alcock and B.~Paczynski, {\it {An evolution free test for non-zero
  cosmological constant}},  {\em Nature} {\bf 281} (1979) 358--359.

\bibitem{Lopez-Corredoira:2013lca}
M.~Lopez-Corredoira, {\it {Alcock-Paczynski cosmological test}},  {\em
  Astrophys. J.} {\bf 781} (2014), no.~2 96,
  [\href{http://arxiv.org/abs/1312.0003}{{\tt arXiv:1312.0003}}].

\bibitem{Sutter:2012tf}
P.~M. Sutter, G.~Lavaux, B.~D. Wandelt, and D.~H. Weinberg, {\it {A first
  application of the Alcock-Paczynski test to stacked cosmic voids}},  {\em
  Astrophys. J.} {\bf 761} (2012) 187,
  [\href{http://arxiv.org/abs/1208.1058}{{\tt arXiv:1208.1058}}].

\bibitem{Blake:2011ep}
C.~Blake et~al., {\it {The WiggleZ Dark Energy Survey: measuring the cosmic
  expansion history using the Alcock-Paczynski test and distant supernovae}},
  {\em Mon. Not. Roy. Astron. Soc.} {\bf 418} (2011) 1725--1735,
  [\href{http://arxiv.org/abs/1108.2637}{{\tt arXiv:1108.2637}}].

\bibitem{Ross:2006me}
N.~P. Ross et~al., {\it {The 2dF-SDSS LRG and QSO Survey: The 2-Point
  Correlation Function and Redshift-Space Distortions}},  {\em Mon. Not. Roy.
  Astron. Soc.} {\bf 381} (2007) 573--588,
  [\href{http://arxiv.org/abs/astro-ph/0612400}{{\tt astro-ph/0612400}}].

\bibitem{Marinoni:2010yoa}
C.~Marinoni and A.~Buzzi, {\it {A geometric measure of dark energy with pairs
  of galaxies}},  {\em Nature} {\bf 468} (2010), no.~7323 539--541.

\bibitem{daAngela:2005gk}
J.~da~Angela, P.~J. Outram, and T.~Shanks, {\it {Constraining beta(z) and Omega
  0(m) from redshift-space distortions in z~3 galaxy surveys}},  {\em Mon. Not.
  Roy. Astron. Soc.} {\bf 361} (2005) 879--886,
  [\href{http://arxiv.org/abs/astro-ph/0505469}{{\tt astro-ph/0505469}}].

\bibitem{Outram:2003ew}
P.~J. Outram, T.~Shanks, B.~J. Boyle, S.~M. Croom, F.~Hoyle, N.~S. Loaring,
  L.~Miller, and R.~J. Smith, {\it {The 2df qso redshift survey. 13. A
  measurement of lambda from the qso power spectrum}},  {\em Mon. Not. Roy.
  Astron. Soc.} {\bf 348} (2004) 745,
  [\href{http://arxiv.org/abs/astro-ph/0310873}{{\tt astro-ph/0310873}}].

\bibitem{Anderson:2012sa}
L.~Anderson et~al., {\it {The clustering of galaxies in the SDSS-III Baryon
  Oscillation Spectroscopic Survey: Baryon Acoustic Oscillations in the Data
  Release 9 Spectroscopic Galaxy Sample}},  {\em Mon. Not. Roy. Astron. Soc.}
  {\bf 427} (2013), no.~4 3435--3467,
  [\href{http://arxiv.org/abs/1203.6594}{{\tt arXiv:1203.6594}}].

\bibitem{Paris:2012iw}
I.~Paris et~al., {\it {The Sloan Digital Sky Survey quasar catalog: ninth data
  release}},  {\em Astron. Astrophys.} {\bf 548} (2012) A66,
  [\href{http://arxiv.org/abs/1210.5166}{{\tt arXiv:1210.5166}}].

\bibitem{Schneider:2010hm}
{\bf SDSS} Collaboration, D.~P. Schneider et~al., {\it {The Sloan Digital Sky
  Survey Quasar Catalog V. Seventh Data Release}},  {\em Astron. J.} {\bf 139}
  (2010) 2360--2373, [\href{http://arxiv.org/abs/1004.1167}{{\tt
  arXiv:1004.1167}}].

\bibitem{Simon:2004tf}
J.~Simon, L.~Verde, and R.~Jimenez, {\it {Constraints on the redshift
  dependence of the dark energy potential}},  {\em Phys. Rev.} {\bf D71} (2005)
  123001, [\href{http://arxiv.org/abs/astro-ph/0412269}{{\tt
  astro-ph/0412269}}].

\bibitem{Stern:2009ep}
D.~Stern, R.~Jimenez, L.~Verde, M.~Kamionkowski, and S.~A. Stanford, {\it
  {Cosmic Chronometers: Constraining the Equation of State of Dark Energy. I:
  H(z) Measurements}},  {\em JCAP} {\bf 1002} (2010) 008,
  [\href{http://arxiv.org/abs/0907.3149}{{\tt arXiv:0907.3149}}].

\bibitem{Moresco:2012jh}
M.~Moresco et~al., {\it {Improved constraints on the expansion rate of the
  Universe up to z~1.1 from the spectroscopic evolution of cosmic
  chronometers}},  {\em JCAP} {\bf 1208} (2012) 006,
  [\href{http://arxiv.org/abs/1201.3609}{{\tt arXiv:1201.3609}}].

\bibitem{Hastings:1970aa}
W.~K. Hastings, {\it {Monte Carlo Sampling Methods Using Markov Chains and
  Their Applications}},  {\em Biometrika} {\bf 57} (1970) 97--109.

\bibitem{Metropolis:1953am}
N.~Metropolis, A.~W. Rosenbluth, M.~N. Rosenbluth, A.~H. Teller, and E.~Teller,
  {\it {Equation of state calculations by fast computing machines}},  {\em J.
  Chem. Phys.} {\bf 21} (1953) 1087--1092.

\bibitem{Hu:1995en}
W.~Hu and N.~Sugiyama, {\it {Small scale cosmological perturbations: An
  analytic approach}},  {\em Astrophys. J.} {\bf 471} (1996) 542--570,
  [\href{http://arxiv.org/abs/astro-ph/9510117}{{\tt astro-ph/9510117}}].

\bibitem{Ade:2015rim}
{\bf Planck} Collaboration, P.~A.~R. Ade et~al., {\it {Planck 2015 results.
  XIV. Dark energy and modified gravity}},
  \href{http://arxiv.org/abs/1502.01590}{{\tt arXiv:1502.01590}}.

\end{thebibliography}

\providecommand{\href}[2]{#2}\begingroup\raggedright\endgroup

\end{document}